\documentclass[twocolumn, 12pt]{aastex631}

\usepackage{placeins}  % \usepackage[section]{placeins}
\usepackage{afterpage}
\usepackage{lipsum}
\usepackage{booktabs}
\usepackage{amsmath}
\usepackage{flafter}
\usepackage{xcolor}
\usepackage{graphicx}
\usepackage{xspace}
\usepackage{threeparttable}

\makeatletter
\appdef\@mkpream@relax{\let\@endpbox\relax\let\do@row@strut\relax\let\ar@align@mcell\relax}
\makeatother

\makeatletter
\def\ApjSectionpenalty{\@M}
\makeatother

\newcommand{\vsini}{\emph{v}$\sin$\emph{i}$_{\ast}$\xspace}

\newcommand{\istar}{\emph{i}$_{\ast}$\xspace}
\newcommand{\idisk}{\emph{i}$_{\mathrm{disk}}$\xspace}
\newcommand{\iorb}{\emph{i}$_{\mathrm{orb}}$\xspace}
\newcommand{\Rstar}{\emph{R}$_{\ast}$\xspace}
\newcommand{\Rsun}{\emph{R}$_{\odot}$\xspace}

\newcommand{\Rearth}{\emph{R}$_{\oplus}$\xspace}
\newcommand{\Mstar}{\emph{M}$_{\ast}$\xspace}
\newcommand{\Msun}{\emph{M}$_{\odot}$\xspace}

\newcommand{\Teff}{\emph{T}$_{\mathrm{eff}}$\xspace}
\newcommand{\Prot}{\emph{P}$_{\mathrm{rot}}$\xspace}
\newcommand{\Deltai}{$\Delta$\emph{i}\xspace}

\newcommand{\TESS}{\textit{TESS}\xspace}
\newcommand{\ktwo}{\textit{K2}\xspace}

\newcommand{\nddnum}{24}
\newcommand{\nwjnum}{21}
\newcommand{\nhjnum}{20}
\newcommand{\nppd}{49\xspace}
\newcommand{\ndd}{24\xspace}
\newcommand{\nwj}{21\xspace}
\newcommand{\nhj}{20\xspace}
\newcommand{\ntot}{114\xspace}

\newcommand{\sigmaidisk}{$\sigma_{i_{\mathrm{disk}}}$\xspace}

\newcommand{\MisProbPDTen}{$33^{+7}_{-6}$\%\xspace}

\newcommand{\MisProbDDTen}{$21^{+9}_{-7}$\%\xspace}

\newcommand{\MisProbWJTen}{$23^{+10}_{-8}$\%\xspace}

\newcommand{\MisProbHJTen}{$20^{+9}_{-8}$\%\xspace}

\shorttitle{Mapping Stellar Obliquity Across Space and Time}
\shortauthors{Biddle \& Bowler}

\begin{document}

\title{Mapping Stellar Obliquity Across Space and Time}

\author[0000-0003-2646-3727]{Lauren I. Biddle}
\affiliation{Department of Astrophysical Sciences, Princeton University, 4 Ivy Lane, Princeton, NJ 08540, USA}
\email{lbiddle@utexas.edu}

\author[0000-0003-2649-2288]{Brendan P. Bowler}
\affiliation{Department of Physics, University of California, Santa Barbara, CA 93106, USA}

\begin{abstract}
The angle between a star’s rotational angular momentum vector and the orbital angular momentum vector of its planets---stellar obliquity---encodes the initial conditions and dynamical history of planetary systems. Variations in obliquity with orbital separation may arise from either primordial disk structure (e.g., broken or warped protoplanetary disks with misaligned inner and outer regions) or subsequent dynamical interactions with planets or external perturbers. Comparing obliquities across orbital distance therefore probes the spatial imprint of these processes, while comparisons across age constrain their temporal evolution. To test for signatures of obliquity evolution in both space and time, we compare stellar inclinations with orbital inclinations for \ntot single Sun-like stars spanning four samples that probe distinct regimes of orbital separation and age: \nppd protoplanetary disk systems, \ndd debris disk systems, \nwj warm Jupiter systems, and \nhj hot Jupiter systems. We find minimum misalignment rates of \MisProbPDTen for protoplanetary disk systems, \MisProbDDTen for debris disk systems, \MisProbWJTen for warm Jupiter systems, and \MisProbHJTen for hot Jupiter systems. These rates are consistent within 1$\sigma$ across all samples. The absence of significant variation in minimum-obliquity distributions with either orbital separation or system age suggests that stellar obliquities for Sun-like stars may be largely established at formation with limited subsequent evolution in both space and time. These results are consistent with post-formation inclination evolution being either infrequent, confined to small misalignments, or restricted to regions of parameter space not examined here---such as systems with close stellar companions, hot host stars, small planets, and long-period giant planets.

\end{abstract}


% ----------- INTRODUCTION ----------- %
\section{Introduction} \label{sec:Introduction}

A fundamental characteristic of planetary system geometry is the stellar spin-orbit obliquity---the angle between a star’s rotational angular momentum vector and the orbital angular momentum vector of its planets. Stellar obliquity is an outcome of the processes that shape planetary systems, and can provide information about past dynamical encounters such as planet-planet scattering events (e.g., \citealt{McArthur2010, Mills2017, BardalezGagliuffi2025}) and migration \citep{Dawson2018, Albrecht2022}. A star’s alignment or misalignment with planetary orbits is often measured using the Rossiter-McLaughlin eﬀect \citep{Rossiter1924, McLaughlin1924} or Doppler shadowing \citep{Albrecht2007, CollierCameron2010} in systems with transiting planets. These methods comprise the majority of measured stellar obliquities to date \citep{Albrecht2022, Southworth2026}, and are most sensitive to giant planets with orbital distances $\lesssim$0.1~au because of the requirement for transiting geometry and the dependence on planet size. A lower limit on the spin-orbit obliquity can be placed based on the star's inclination, determined from its radius, rotational period, and projected rotational velocity (e.g., \citealt{Hale1994, Masuda2020, Bowler2023}). This approach is especially useful for distant transiting planets and non-transiting systems, for instance, using absolute astrometry, for directly imaged planets, and for resolved circumstellar disks. 

As a result, the distribution in stellar obliquities for transiting giant planets is well-established, revealing a broad range in stellar obliquity angles from aligned configurations (0$^{\circ}$) to retrograde ($>$90$^{\circ}$) orbits \citep{Winn2015, Albrecht2022}. A trend emerged early on between a system's spin-orbit obliquity and stellar effective temperature: the orbits of hot Jupiters ($\lesssim$0.1~au) are often more misaligned around hot stars ($\gtrsim$6200~K) compared to cool stars \citep{Winn2010, Schlaufman2010}. A leading explanation for the preferential alignment of cool stars is that they have large convective envelopes compared to hot stars \citep{Pinsonneault2001}, which allow tidal forces to align the star (e.g., \citealt{Lai2012}) from an initially misaligned configuration that is otherwise preserved in hot star systems \citep{Winn2010, Schlaufman2010, Albrecht2012}.

Warm Jupiters ($\lesssim$0.1--1~au) are less sensitive to the effects of tidal forces \citep{Dawson2014a}, and so spin orbit obliquities of warm Jupiter host stars better reflect initial system geometries and facilitate probes of primordial (mis)alignment. As a population, warm Jupiter systems tend to exhibit aligned configurations \citep{Rice2022a, Morgan2024, Wang2024}, implying aligned primordial spin-orbit obliquities. Most transiting planet systems, however, have been subject to Gyrs of post-formation dynamics. Young systems are more suitable to address the evolution of stellar obliquity, but strong intrinsic variability in stellar spectra and photometry make planets difficult to detect and prevent detailed characterization of their obliquities. As a result, few obliquities have been measured in systems $<$100~Myr \citep{Albrecht2022}\footnote{One exception is for directly imaged planets, which typically orbit young ($\lesssim$200~Myr) stars at wide separations (5-100~au). \citet{Bowler2023} found that young giant planets are preferentially aligned with their host stars, but the sample of imaged planets with well-constrained orbits and stellar inclinations remains modest \citep{Sepulveda2024a, Balmer2026}.}.
 
Debris disks---second generation circumstellar disks that span young and old ages (tens of Myr to a few Gyr)---offer another way to examine how primordial obliquities evolve. \citet{Hurt2023} compared debris disk ($\gtrsim$10~au) inclinations derived from resolved imaging to stellar inclinations inferred from rotation periods, radii, and projected rotational velocities to determine lower limit on star-disk obliquities. Among 31 systems, six were significantly misaligned; the majority were consistent with alignment.

Recently, an analysis of 49 young ($\sim$2–5~Myr) single Sun-like stars with resolved protoplanetary disks provided a direct characterization of primordial spin-orbit obliquities \citep{Biddle2025}, benchmarking the earliest configurations of planetary systems. The reconstructed distribution of minimum star-disk obliquities revealed a broad distribution reaching out to about 50$^{\circ}$. The majority of systems were found to be consistent with spin-orbit alignment, while one-third were misaligned by at least 10$^{\circ}$. This implies that most planets form out of aligned disks, although a significant fraction of systems may be modestly misaligned at the epoch of planet formation.

Among stars harboring hot Jupiters, warm Jupiters, debris disks, and protoplanetary disks, spin-orbit obliquities have now been sampled across wide ranges of age and orbital separation. Comparing obliquities of populations at different ages and separations can help gauge the impact of post-formation dynamics on spin-orbit geometries over time. Changes in obliquity that occur after the epoch of planet-formation would result from dynamical interactions that excite planetary orbital inclinations, and would manifest in observations as a departure from the primordial distribution.

Several mechanisms may alter stellar obliquities over time. For example, dynamical interactions among multiple planets can excite mutual inclinations and misalignment \citep{Chatterjee2008, Juric2008, Chatterjee2011, Mills2017}. Similarly, an inclined outer companion can produce large-amplitude exchanges between eccentricity and inclination via von Zeipel–Kozai-Lidov (ZKL oscillations \citep{Fabrycky2007, Storch2014}. Close encounters from passing stars may also tilt planetary systems \citep{Malmberg2011, Li2015}. Moreover, the orientations of stellar spin axes themselves may evolve over long timescales through angular-momentum evolution \citep{Lai2011, Spalding2014}.

Here, we assess whether the primordial obliquity distribution found in young protoplanetary disk systems is preserved in intermediate-age debris disks, older warm Jupiter systems, and mature hot Jupiter systems. We draw our samples from \citet{Biddle2025}, \citet{Hurt2023}, and \citet{Morgan2024} with careful updates to each sample to ensure the hosts are Sun-like stars with similar characteristics. Each sample represents a distinct spatial and temporal window into planetary angular momentum architectures around cool stars (0.3~$\lesssim$~\Mstar~$\lesssim$~1.2~\Msun, 3100~$\lesssim$~\Teff~$\lesssim$6200~K), with which we assess and compare snapshots of their spin-orbit obliquities. The samples and relevant observations are detailed in Section \ref{sec:Sample_Descriptions}, and the statistical methods and analyses are described in Section \ref{sec:Results}. Section \ref{sec:Discussion} provides discussion of the results and the implications for obliquity evolution. Section \ref{sec:Summary} concludes with a summarization of this work.

% --------------- OBSERVATIONS --------------- %
\section{Sample Descriptions}\label{sec:Sample_Descriptions}

Systems included in this work consist of stars with measured projected rotational velocities (\vsini) and rotation periods (\Prot), which can be used to determine the stellar inclination (\istar). We select stars with no known stellar or substellar companions within 10,000~au, and omit stars with effective temperatures (\Teff) greater than 6200~K to avoid $\gamma$~Dor- and $\delta$~Scuti-like variables \citep{Kaye1999, Aerts2021} which can produce periodic brightness signatures that have the potential to be falsely attributed to rotation (e.g., \citealt{Sepulveda2022}). The full sample consists of \ntot stars split into four subsamples: \nppd stars that host a protoplanetary disk, \ndd stars that host a debris disk, \nwj stars that host a transiting warm Jupiter, and \nhj stars that host a transiting hot Jupiter. More information about the sample selection is detailed below. Adopted stellar masses, effective temperatures, radii, rotation periods, and projected rotational velocities for all stars included in this work are provided in Table \ref{tab:Sample_Properties} in Appendix \ref{appendix:Tables}.

% --- PROTOPLANETARY DISKS --- %
\subsection{Protoplanetary Disk Sample}\label{subsec:PPD_Sample}

The subsample of protoplanetary disk-bearing stars is adopted from \citet{Biddle2025}. This sample consists of \nppd cool (\Teff$<$6200~K) T~Tauri stars, primarily in the Taurus, Lupus, and $\rho$-Ophiuchus star-forming regions, that host protoplanetary disks with submillimeter dust emission that has been spatially resolved with the Atacama Large Millimeter/submillimeter Array (ALMA). The stars are gravitationally isolated with no known stellar companions, and reside in low-density regions where past stellar fly-bys are unlikely to have occurred \citep{Winter2024}. Disk orientations derived from ALMA interferometry trace dust emission in the outer regions of the disk, typically spanning separations of tens to hundreds of au. Inclinations of the disks (\idisk) in this population are adopted from the literature, and include only measurements with \sigmaidisk$<$6$^{\circ}$. On average, \sigmaidisk=2.6$^{\circ}$. The stellar properties (\vsini, \Rstar, and \Prot) used to determine stellar inclinations (\istar) were derived from the weighted mean of compiled values from the literature with the addition of new \Prot measurements from homogeneously analyzed \TESS and \ktwo light curves where applicable. Across the sample, stellar masses and effective temperatures fall primarily between $\sim$0.3--1.2~\Msun and $\sim$3100--5600~K, respectively.

% --- DEBRIS DISKS --- %
\subsection{Debris Disk Sample}\label{subsec:DD_Sample}

The sample of debris disk systems originates from the catalog constructed by \citet{Hurt2023}, which consists of disks that have been spatially resolved in far-infrared to millimeter wavelength observations with the \textit{Herschel Space Observatory} \citep{Pilbratt2010}, Gemini Planet Imager \citep{Macintosh2014}, SPHERE/VLT \citep{Beuzit2008, Fusco2014}, and ALMA. Here, \idisk describes the orientation of the outer debris disk, tracing radial separations that are broadly similar to protoplanetary disks (tens to hundreds of AU; \citealt{Pearce2022}). Each system also has measurements of stellar \vsini, \Rstar, \Prot to determine the inclination of the central star.

Out of 31 systems in the original catalog, we retain stars with \Teff$<$6200~K to preserve consistency in the stellar demographics between populations. Several debris disk systems have new \idisk measurements with improved precision (see Table \ref{tab:DD_Disk_Inclinations}), which we adopt in this work. For consistency with the selection criteria for protoplanetary disks, we keep 18 systems with $\sigma_{i_{\mathrm{disk}}}$$<$6$^{\circ}$. From this list, we remove one system with no reported \idisk measurement uncertainty (HD~377), and one system that has recently been identified as a spectroscopic binary (HD~129590; \citealt{Grandjean2023}). To the 16 stars remaining from the original catalog, we add six systems with recent disk observations (BD+45~598, Fomalhaut~C, GJ~14, HD~105, HD~207129, and HD~38858; \citealt{Cronin-Coltsmann2021, Marshall2021, Vincent2024, Matra2025}) for a final sample size of \nddnum. Fomalhaut~C resides in a wide triple system with a separation $>$10,000~au \citep{Mamajek2013} to Fomalhaut~AB. We retain it in the sample, as companions at these separations are not expected to drive significant dynamical evolution \citep{Holman1999, Trilling2007, Moe2021}. The catalog of debris disk systems analyzed in this work is listed in Table \ref{tab:Sample_Properties}.

To ensure consistency in the adoption of system parameters across every sample, we re-compute \vsini and \Prot in the debris disk sample using methods identical to that of the other samples \citep{Morgan2024, Biddle2025}. In the original \citet{Hurt2023} catalog, the stellar \vsini measurements were selected from the literature based on the highest-resolution spectrum, and \Prot measurements were derived from homogeneously analyzed \TESS light curves using Gaussian Process regression. Rather than rely on a single reported value, we update the adopted \vsini values so that they are determined from the weighted mean of published measurements, and impose a minimum uncertainty of 5\% to mitigate for systematics in the measurements and underestimated uncertainties. 

We also re-analyze \TESS light curves using the same Lomb-Scargle period search routine \citep{VanderPlas2018} as was carried out for the protoplanetary disk \citep{Biddle2025} and transiting planet samples \citep{Morgan2024}. From the resulting periodograms, we take the \Prot to be equal to the periodic signal associated with the highest peak in the power spectrum and adopt the Half Width at Half Maximum of that peak as the uncertainty. We then inflate the uncertainty to account for possible effects from differential rotation, which can result in an over estimation of \Prot. We assume a Sun-like absolute shear of 0.07~rad~d$^{-1}$ and add the error term associated with this shear to the nominal \Prot uncertainty in quadrature (e.g., \citealt{Bowler2023}). This increases the uncertainties of the half-width-half-max estimation by $\sim$70\%, on average. The original values of \Rstar from \citet{Hurt2023} are adopted from the TESS Input Catalog (TIC; \citealt{Stassun2019}). This methodology agrees with the other samples; however, we amend the uncertainties by adding a 7\% error term in quadrature to the quoted uncertainty in the TIC, which accounts for underestimated stellar radius estimates based on a comparison with asteroseismic measurements \citep{Huber2017}. The compilation of \Rstar, \Prot, and \vsini is provided in Table \ref{tab:DD_Rstar_Prot_vsini} in Appendix \ref{appendix:Tables}. Adopted \idisk measurements are given in Table \ref{tab:DD_Disk_Inclinations} in Appendix \ref{appendix:Tables}.

% --- WARM AND HOT JUPITERS --- %
\subsection{Planet Host Samples}\label{subsec:J_Sample}
Both subsamples of stars with transiting giant planets originate from the compilation by \citet{Morgan2024}, who report lower limits on the spin-orbit obliquities of 36 hot and 24 warm Jupiter systems with a minimum planetary mass ($m\sin i_{\ast}$) between 0.3--13~$M_{\mathrm{Jup}}$ or planetary radius ($R_{\mathrm{p}}$) $>$8~\Rearth. Here, we adopt their definition of warm and hot Jupiters, where warm Jupiters reside at a scaled orbital distance (\emph{a}/\Rstar) $\geq$20 (corresponding to a semimajor axis of $\gtrsim$1~au), and hot Jupiters orbit at a scaled distance \emph{a}/\Rstar$<$20 ($\lesssim$1~au for a Sun-like star). They found that the majority of both hot and warm Jupiter host stars agree with an aligned orientation with similar misalignment rates ($\sim$14--24\%). 

The method used to compile system parameters of the warm and hot Jupiter samples is the same as for the protoplanetary and debris disk samples. The primary difference is that the inclinations of transiting planets (\iorb) are viewed edge-on, so \iorb$\approx$90$^{\circ}$ for all systems. \citet{Morgan2024} adopted the weighted mean of \vsini and \Rstar values in the literature, and the stellar rotation periods are derived from a homogeneous analysis of \TESS, \emph{Kepler}, and \ktwo lightcurves (see Tables~1 and 4 from \citealt{Morgan2024}). From the original catalog, we select only systems that are single stars or separated by a companion star by $>$10,000~au, resulting in \nwjnum\, systems with warm Jupiters and \nhjnum\, systems with hot Jupiters. 

\begin{figure*}[ht!]
    \centering
    \includegraphics[width=\textwidth]{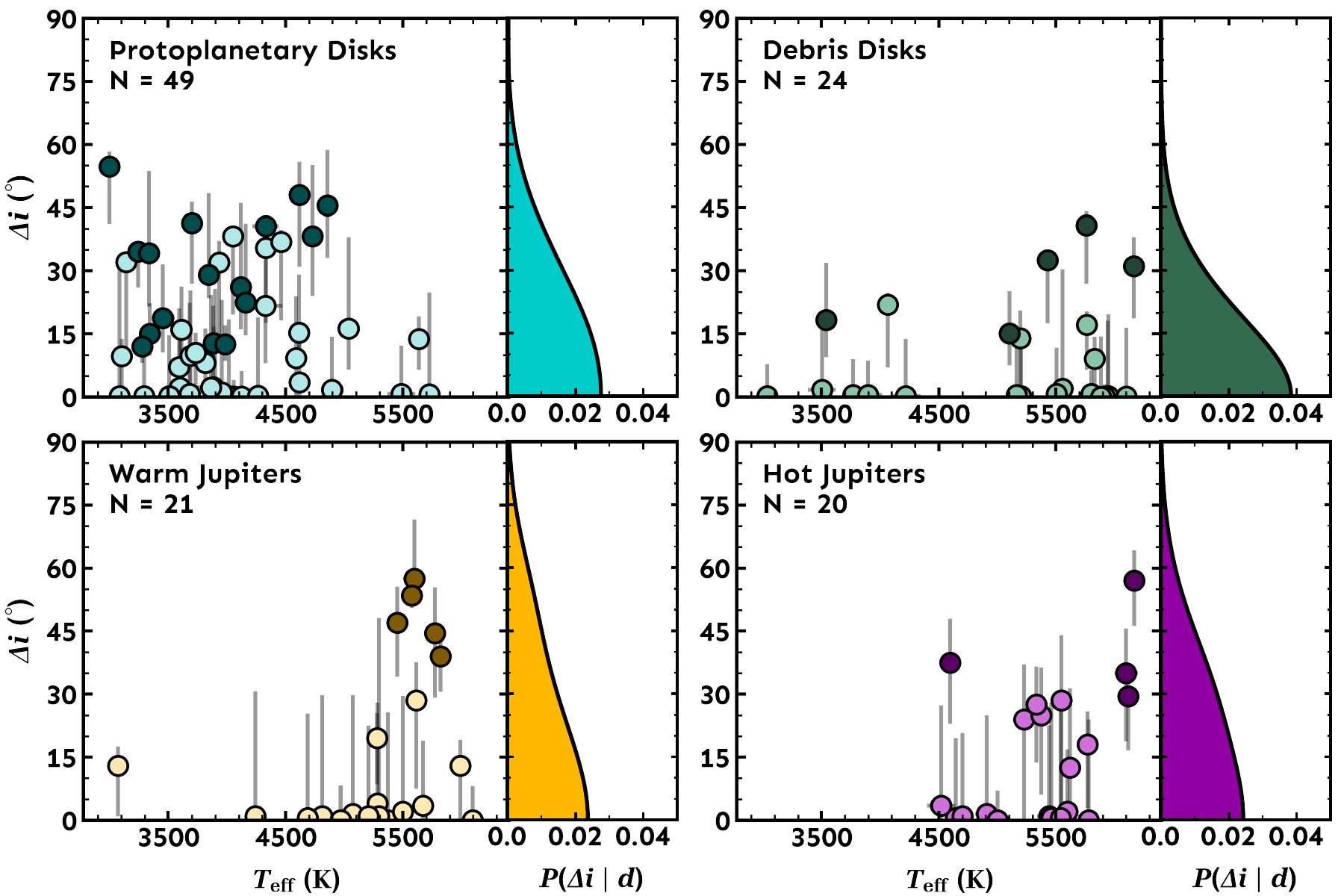}
    \caption{Minimum obliquities (\Deltai) within each sample as a function of stellar \Teff, shown alongside their respective KDE. Dark colored points represent systems that are classified as misaligned (Section \ref{sec:Test2}), and light colored points are systems that are consistent with alignment. \textit{N}~is the size of the sample.}
	\label{fig:Deltai}
\end{figure*}

% ---- STELLAR INCLINATIONS ---- %
\subsection{Stellar Inclinations}\label{subsec:Stellar_Inclinations}

Stellar inclinations are uniformly re-derived within each sample to determine the probability distribution of minimum obliquity, \Deltai. We use \Rstar, \vsini, and \Prot to infer the posterior distribution of \istar for each star using the analytical expression developed by \citet{Bowler2023}. The expression follows the Bayesian probabilistic framework presented by \citet{Masuda2020}, and assumes an isotropic prior on \istar with uniform priors on \vsini, \Rstar, and \Prot:

\begin{equation}\label{eq:istar}
P(i_* \mid P_\mathrm{rot}, R_{*}, v\sin i_{*}) \propto  \sin i_* \times \frac{e^{- \frac{\big(v \sin i_* - \frac{2\pi R_*}{P_\mathrm{rot}}\sin i_* \big)^2}{2\big(\sigma_{v\sin i_*}^2 + \sigma_{v_\mathrm{eq}}^2 \sin^2 i_* \big)}} }{\sqrt{\sigma_{v\sin i_*}^2 + \sigma_{v_\mathrm{eq}}^2 \sin^2 i_*}},
\end{equation}
where
\begin{equation}\label{eq:veq_err}
\sigma_{v_\mathrm{eq}} =  \frac{2\pi R_{*}}{P_\mathrm{rot}} \sqrt{\Big(\frac{\sigma_{R_{*}}}{R_{*}}\Big)^2 + \Big(\frac{\sigma_{P_\mathrm{rot}}}{P_\mathrm{rot}}\Big)^2 },
\end{equation}
with the uncertainties on the rotation period, projected rotational velocity, and stellar radius given by $\sigma_{P_\mathrm{rot}}$, $\sigma_{v \sin i_{*}}$, and $\sigma_{R_{*}}$, respectively. This expression can be modified to include an upper limit on vsini, $l$:
\begin{equation}{\label{eqn:i_ul}}
\begin{split}
P(i_* \mid P_\mathrm{rot}, R_{*}, v\sin i_{*}) \propto \sin i_* \times& \Big( \mathrm{erf}\Big(\frac{l - \frac{2\pi R_*}{P_\mathrm{rot}} \sin i_*}{\sqrt{2} \, \sigma_{v_\mathrm{eq}} \sin i_*}  \Big) \\
    & + \mathrm{erf}\Big(\frac{\sqrt{2} \, \pi R_*}{\sigma_{v_\mathrm{eq}} P_\mathrm{rot} }  \Big)   \Big).
\end{split}
\end{equation}
We use Equation \ref{eqn:i_ul} for the protoplanetary disk-bearing stars: 2MASS~J16083070-3828268, GW~Lup, Sz~114, and Sz~90, for the debris disk star, 61~Vir, and for the warm Jupiter host, Kepler-1654. Our reported measurements and uncertainties of \istar are derived from maximum \textit{a posteriori} (MAP) value and 68\% highest density interval of the posterior distributions, and are listed in Table \ref{tab:Sample_Properties} in Appendix \ref{appendix:Tables}. The \istar posterior distributions of protoplanetary disk stars can be found in \citet{Biddle2025}. We display the individual distributions of debris disk stars in Figure \ref{fig:sample_posteriors_1} in Appendix \ref{appendix:Distributions}. Tabulated warm and hot Jupiter host star inclinations can be found in \citet{Morgan2024} and in Table \ref{tab:Sample_Properties}.

% --------------- RESULTS--------------- %
\section{Results}\label{sec:Results}

% ---- MINIMUM OBLIQUITY ---- %
\subsection{Minimum Obliquity, $\Delta i$}\label{subsec:Star-Disk_Obliquity}

The true stellar obliquity ($\Psi$) is defined as the angle between angular momentum vectors of the star's rotation ($\overset{\rightharpoonup}L_{*}$) and the orbit of the disk or planet ($\overset{\rightharpoonup}L_{\mathrm{disk}}$ or $\overset{\rightharpoonup}L_{\mathrm{orb}}$), which requires that the full spatial orientations of the star and disk be known. In the case of disk-bearing stars, the orientation of the star's spin is usually unknown, and the kinematic measurements necessary to describe whether $\overset{\rightharpoonup}L_{\mathrm{disk}}$ points toward or away from Earth are not always available. Therefore, to allow direct comparisons between star-disk obliquities and spin-orbit obliquities, we compute the minimum obliquity ($\Delta i$~$=$~$|i_{*}-i_{\mathrm{disk}}|$), which corresponds to the lower limit of $\Psi$, and is equal to the absolute difference between the sky-projected inclinations of the stellar rotation axis and the disk or planetary orbital plane. For transiting planet systems, \iorb$\simeq$90$^{\circ}$, making \Deltai effectively equivalent to 90$^{\circ}$$-$\istar.

To determine \Deltai for protoplanetary and debris disk systems, we draw samples from the posterior of \istar and a Gaussian approximation of the \idisk  distribution (with mean and standard deviation set by the adopted value and uncertainty). \Deltai values are computed as the absolute difference for each draw, yielding its distribution function. In Table \ref{tab:Sample_Properties}, MAP values are reported along with uncertainties from the 68\% highest density interval.

% ---- MINIMUM OBLIQUITY ---- %
\subsection{Population-Level Distribution of $\Delta i$}\label{subsec:Deltai_Population_Level}

For each sample, the distribution of \Deltai is characterized using a Kernel Density Estimate (KDE) of the resulting probability distributions of \Deltai. Figure \ref{fig:Deltai} summarizes the individual values and overall distribution of $\Delta i$ in each sample. To fairly compare the distributions, each KDE is constructed using the same broadening parameter ($\kappa$). We choose $\kappa$~=~12.4$^{\circ}$, which is the characteristic precision of the sample with the largest average uncertainty on \Deltai (hot Jupiters).

\begin{figure}[ht!]
    \centering
    \includegraphics[width=\columnwidth]{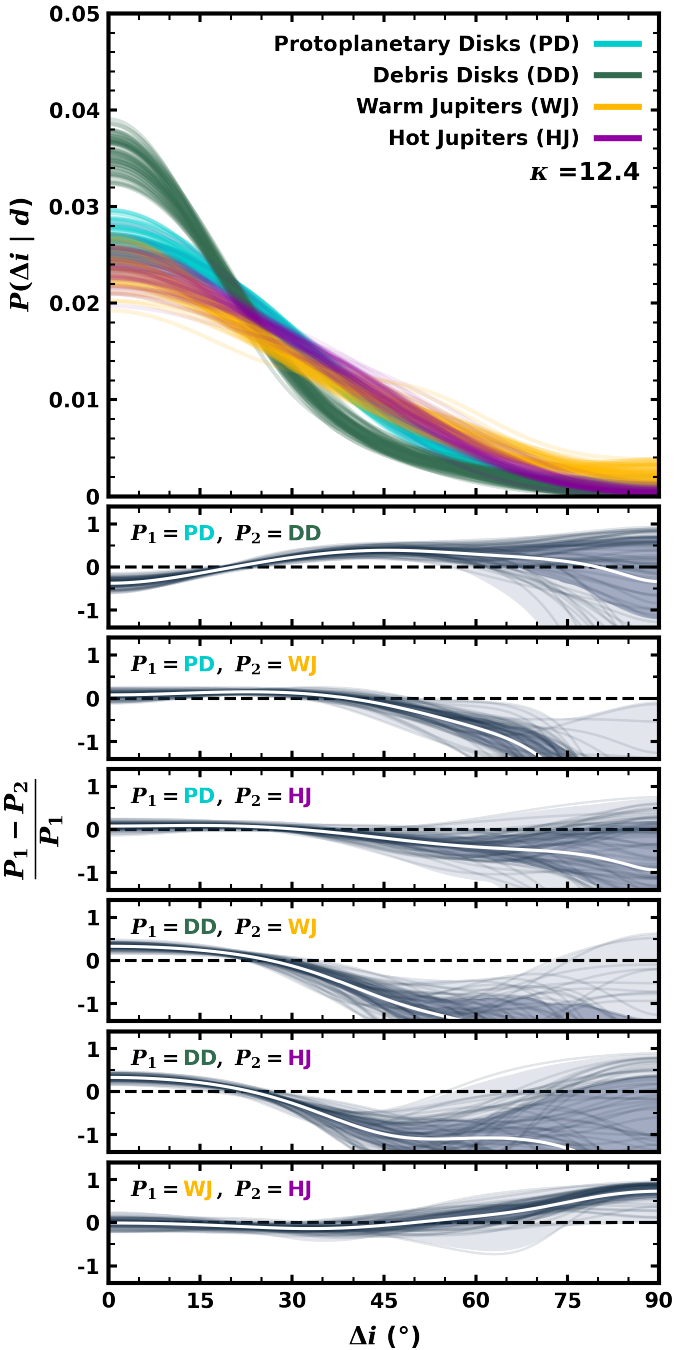}
    \caption{\textit{Panel A}: Family of KDEs that characterize the spread in the \Deltai distribution of each sample. The top panel overlays the family of KDEs for comparison, which are color coded according to the sample. $\kappa$ is the band width used to compute the KDEs. \textit{Panels B–G}: The spread of the residuals of unique pairs of KDEs from two different families ($P_1$ and $P_2$) represented as a fraction of $P_1$. The 68\% (1$\sigma$) and 99.7\% (3$\sigma$) highest density intervals are shown in dark gray and light gray, respectively, with the white line representing the median. The specific samples from which $P_1$ and $P_2$ are drawn are labeled in the upper left corner of each panel. Positive values indicate an excess of systems in $P_1$ relative to $P_2$ at that \Deltai.}
	\label{fig:KDEs}
\end{figure}

Following the steps described in \citet{Biddle2025}, a family of KDEs is generated to explore the range of distributions that reflect the uncertainties of the star and disk measurements used to determine \Deltai. These KDEs are derived using values of \idisk, \vsini, \Prot, and \Rstar that are ascertained by randomly drawing from their respective Normal probability distribution. Figure \ref{fig:KDEs} (Panel A) displays the family of KDEs computed for each sample, overlaid for comparison.

\subsection{Comparing $\Delta i$ Distributions}\label{subsec:Comparing_Distributions}
To test whether stellar obliquity evolves across orbital separation and system age, we compare the \Deltai distributions of each population using two complementary approaches that assess both the overall shape of the distributions and the fraction of systems that are significantly misaligned:

\subsubsection{Test 1: Relative \Deltai Probability Density}\label{sec:Test1}
The relative power of each sample's probability density distribution as a function of \Deltai is used to identify whether one sample is more or less aligned compared to another sample. This comparison is carried out by computing the difference between two \Deltai probability density distributions, $P_1$ and $P_2$ (depicted by KDEs), with the residuals represented as a fraction of $P_1$. To account for uncertainties in the measurements used to determine \Deltai, we repeat this calculation for all distributions in the relevant family of KDEs constructed for each sample (see Section \ref{subsec:Deltai_Population_Level}), resulting in 10,000 sets of residuals for each two-sample comparison.

\begin{figure*}[ht!]
    \centering
    \includegraphics[width=\textwidth]{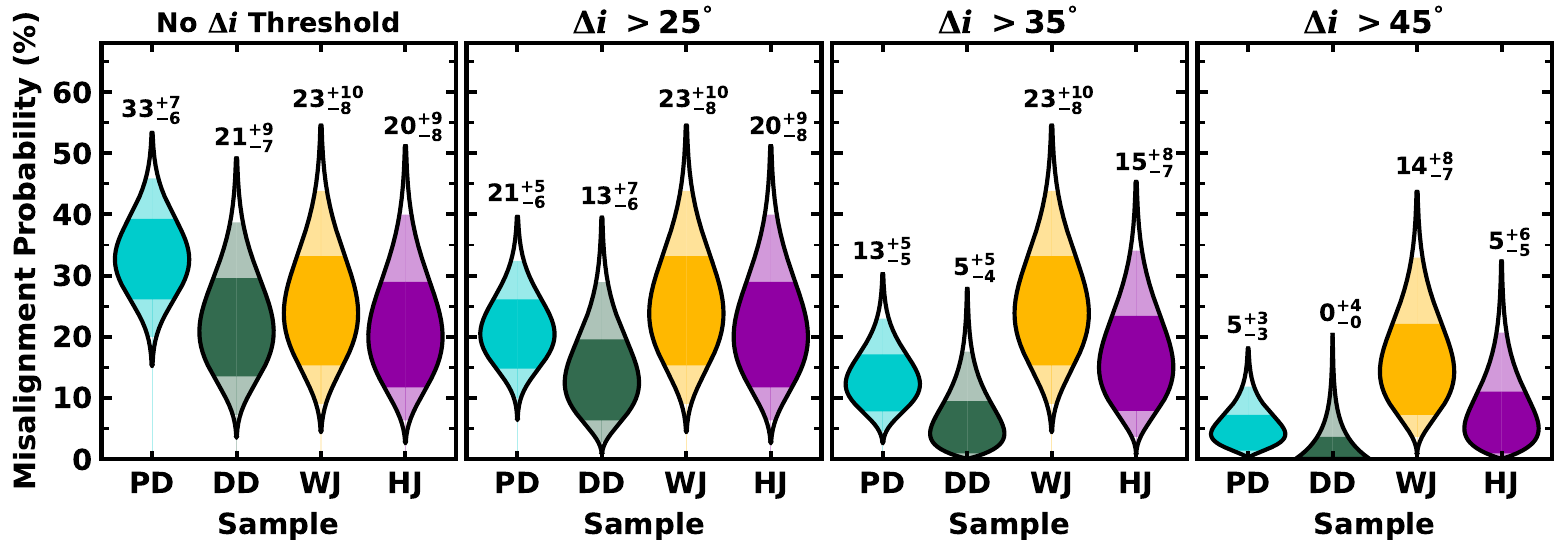}
    \caption{The misalignment probability in each population. The protoplanetary disk population is abbreviated as ``PD'', debris disks as ``DD'', warm Jupiters as ``WJ'', and hot Jupiters as ``HJ''. The leftmost panel shows the misalignment probability distributions with no lower limit on \Deltai when counting misaligned systems. The black lines encapsulate the 3$\sigma$ confidence range. The darker shaded region shows the 1$\sigma$ confidence interval, and the lighter shaded region shows the 2$\sigma$ confidence interval. From center left to right, the misalignment probabilities determined for systems that are misaligned by at least 25$^{\circ}$, 35$^{\circ}$, and 45$^{\circ}$.}
	\label{fig:misalignment_probabilities}
\end{figure*}

Results are shown in Figure \ref{fig:KDEs}. In every set of residuals, the average residual departs from zero by at least 1$\sigma$ over some ranges of \Deltai, and in some cases the difference is significant to $>$3$\sigma$. Among the paired samples, debris disks appear the most discrepant. Relative to the other populations, the debris disk sample exhibits a 30\%--40\% excess in systems with \Deltai~$<$~10$^{\circ}$ and at least a $\sim$45\%--100\% deficit in systems with \Deltai~$\gtrsim$~40$^{\circ}$. In contrast, the \Deltai distributions of protoplanetary disks, warm Jupiters, and hot Jupiters remain generally alike, with relative differences below $\sim$15\% for \Deltai~$\lesssim$~50$^{\circ}$. These qualities are consistent when repeating this analysis using bandwidths that are one-half, and two times, the nominal value. In short, we find that all four samples are broadly similar. There are hints that the debris disk sample may be somewhat more aligned, but, as detailed below, this may relate to the higher precision of the $\Delta i$ measurements for that particular sample.

\subsubsection{Test 2: Misalignment Probability and Magnitude}\label{sec:Test2}
The fraction and degree to which spin-orbit angles are misaligned depends on the specific mechanisms that drive their dynamical evolution. Here, we assess whether the rate of spin-orbit misalignment in each sample changes as a function of separation and age. Following \citet{Biddle2025}, we classify a system as misaligned if the MAP value of its \Deltai probability distribution is greater than 0$^{\circ}$ by at least 2$\sigma$\ ($\geq$95.4\% confidence). For protoplanetary disk systems, the misalignment probability is \MisProbPDTen (16/\nppd; \citealt{Biddle2025}). In the other samples, we find that \MisProbDDTen (5/\ndd) of debris disk systems, \MisProbWJTen (5/\nwj) of warm Jupiter systems, and \MisProbHJTen (4/\nhj) of hot Jupiter systems are misaligned. For all of these, the misalignment probability is consistent within 1$\sigma$. 
    
Next, we assess whether there is a difference in magnitude of \Deltai among the misaligned systems in each sample. The average \Deltai among protoplanetary disks classified as misaligned is 30.5$^{\circ}$. Among debris disks, the average \Deltai values for misaligned systems is 29.4$^{\circ}$, and for warm and hot Jupiter systems the average is 48.4$^{\circ}$ and 39.9$^{\circ}$, respectively. The characteristic precision of \Deltai for the subset of misaligned systems in each sample is similar to that of the full sample. For protoplanetary disks, misaligned \Deltai values have a characteristic precision of 10.1$^{\circ}$. The uncertainties for misaligned debris disk, warm Jupiter, and hot Jupiter systems is 7.8$^{\circ}$, 8.4$^{\circ}$, and 11.3$^{\circ}$, respectively.
    
We next examine the rate of misalignments above increasing \Deltai thresholds. We first measure the rate of misaligned systems with \Deltai $>$10$^{\circ}$, which represents the characteristic uncertainty of \Deltai of all four samples combined (9.9$^{\circ}$). Misaligned systems within all samples have a \Deltai MAP value of at least 10$^{\circ}$. Increasing this threshold by increments of 15$^{\circ}$ up to 55$^{\circ}$, we find that the misalignment rate represented by each sample trends lower, but remains consistent within 1$\sigma$ up to \Deltai$>$40$^{\circ}$. For \Deltai$>$40$^{\circ}$, the misalignment rate between samples is consistent within 2$\sigma$, returning to 1$\sigma$ agreement for \Deltai$>$55$^{\circ}$ (Figure \ref{fig:misalignment_probabilities}). Simply put, we find that in addition to consistency in the average degree of misalignment in each sample, the rate of misalignments above increasing \Deltai thresholds is also similar.

\begin{figure*}[ht!]
    \centering
    \includegraphics[width=\textwidth]{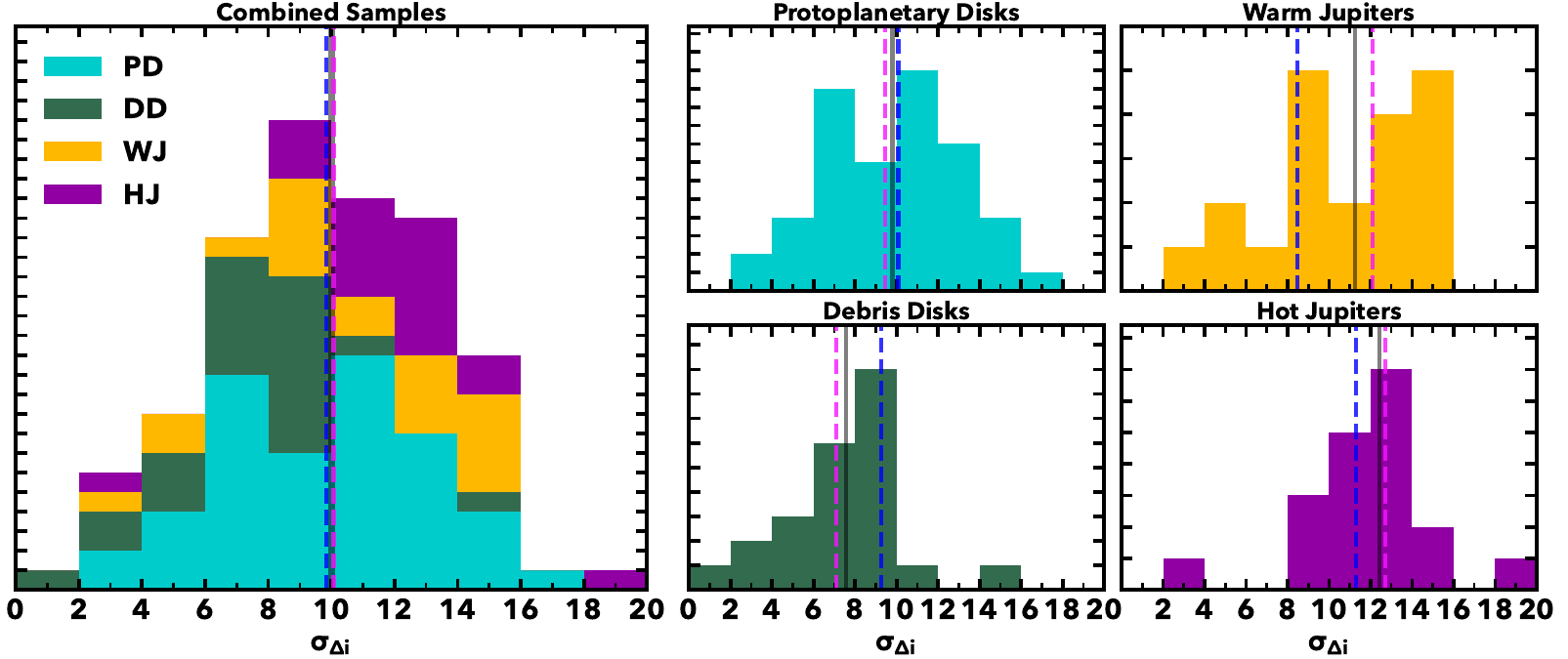}
    \caption{The distributions of the characteristic measurement precision of \Deltai ($\sigma_{\Delta i}$) for all samples combined (left) and individually (center and right). The average $\sigma_{\Delta i}$ of systems that were classified as misaligned is marked by the blue dotted line. For systems that are consistent with alignment, the average $\sigma_{\Delta i}$ is indicated by the magenta dotted line. The gray solid line indicates the average $\sigma_{\Delta i}$ of all systems in the respective panel. Tick marks along the vertical axes represent increasing unit increments of 1.}
    \label{fig:sigmaDi_precision_histogram}
\end{figure*}

% -------------- DISCUSSION -------------- %
\section{Discussion}\label{sec:Discussion}

\subsection{Sample Coverage in Orbital Separation and Age}
The samples examined in this work represent different stages in the evolutionary progression of planetary systems, probing close-in ($\lesssim$0.1~au) and wide ($\gtrsim$10~au) orbital distances. At wide separations ($\sim$10–300~au), the protoplanetary disk sample is the earliest epoch when planets are still forming from the disk (spanning ages between 2--5~Myr old on average; \citealt{Wilking2008, Esplin2020, Kraus2009a, Galli2020, Galli2021}). The sample of debris disks is observed at a similar spatial separations from the central star ($\sim$10–200~au), but covers a later and broader range in ages (4~Myr--4~Gyr; e.g., \citealt{Pizzolato2003, Song2004, Hillenbrand2008, Mentuch2008, Maldonado2010, Chen2011, Bell2015, Aguilera-Gomez2018}) with a median age of 190~Myr. At closer-in separations ($\sim$0.07–2~au), the sample of warm Jupiters ($\sim$20~Myr--12~Gyr) shares some overlap in age with the debris disk sample. However, with a median age of 1.6~Gyr, most warm Jupiters are considerably older, probing a later stage in system development. The sample of hot Jupiter host stars ranges from 2.2--11~Gyr with a median age of $\sim$5.2~Gyr (\citealt{Morgan2024} and references therein).

\subsection{Comparing $\Delta i$ Distributions}
Discrepancies in the probability distributions of \Deltai between samples of chronologically adjacent ages could be indicative of post-formation dynamical evolution. We tested for differences in the probability distributions of \Deltai between pairs of samples and found that debris disks are consistently more likely to have \Deltai closer to 0$^{\circ}$ and less likely to have \Deltai $\gtrsim$30$^{\circ}$ than the other samples (Section \ref{sec:Test1}). Debris disk hosts tend to be more aligned with their star's spin axis than (younger) planet-forming disks and (older) transiting giant planets. It is not clear how a debris disk might reorient an initially misaligned star. Realigning a star through torques from a debris disk is challenging because the stellar angular momentum greatly exceeds that of the low-mass disk, and the absence of gas eliminates the magnetic and accretion torques that operate during the protoplanetary disk phase.

An alternative explanation could be a possible observational bias that favors aligned debris disks over misaligned disks. Such a bias could originate from (for example) quenched stirring in misaligned systems. The detectability of a debris disk largely depends on its ability to continually replenish dust \citep{Wyatt2008} via planet- or self-stirring \citep{Mustill2009, Kennedy2010, Krivov2018, Marshall2026}. So, if misaligned systems are more often subject to some mechanism that causes dynamical heating, planets are more likely to be ejected (e.g., \citealt{Chatterjee2008, Juric2008}), which may disrupt outer planetesimal belts \citep{Raymond2011}.

A simpler explanation for this behavior could be that it is an effect caused by differences in the characteristic precision of \Deltai ($\sigma_{\Delta i}$) in systems with low versus high degrees of misalignment (Figure \ref{fig:sigmaDi_precision_histogram}). Debris disks that are not classified as misaligned (\Deltai consistent with 0$^{\circ}$) have 24\% smaller $\sigma_{\Delta i}$ than those classified as misaligned. This effect is significantly reduced in protoplanetary disks ($\sigma_{\Delta i}$ is only 7\% smaller in non-misaligned stars) and absent entirely from warm and hot Jupiter systems, which instead demonstrate an opposite relationship between $\sigma_{\Delta i}$ and degree of \Deltai (mis)alignment. Increasing the sample sizes and reducing the measurement uncertainties of all populations would help clarify whether debris disks are truly more aligned. 

\begin{figure*}[ht!]
    \centering
    \includegraphics[width=\textwidth]{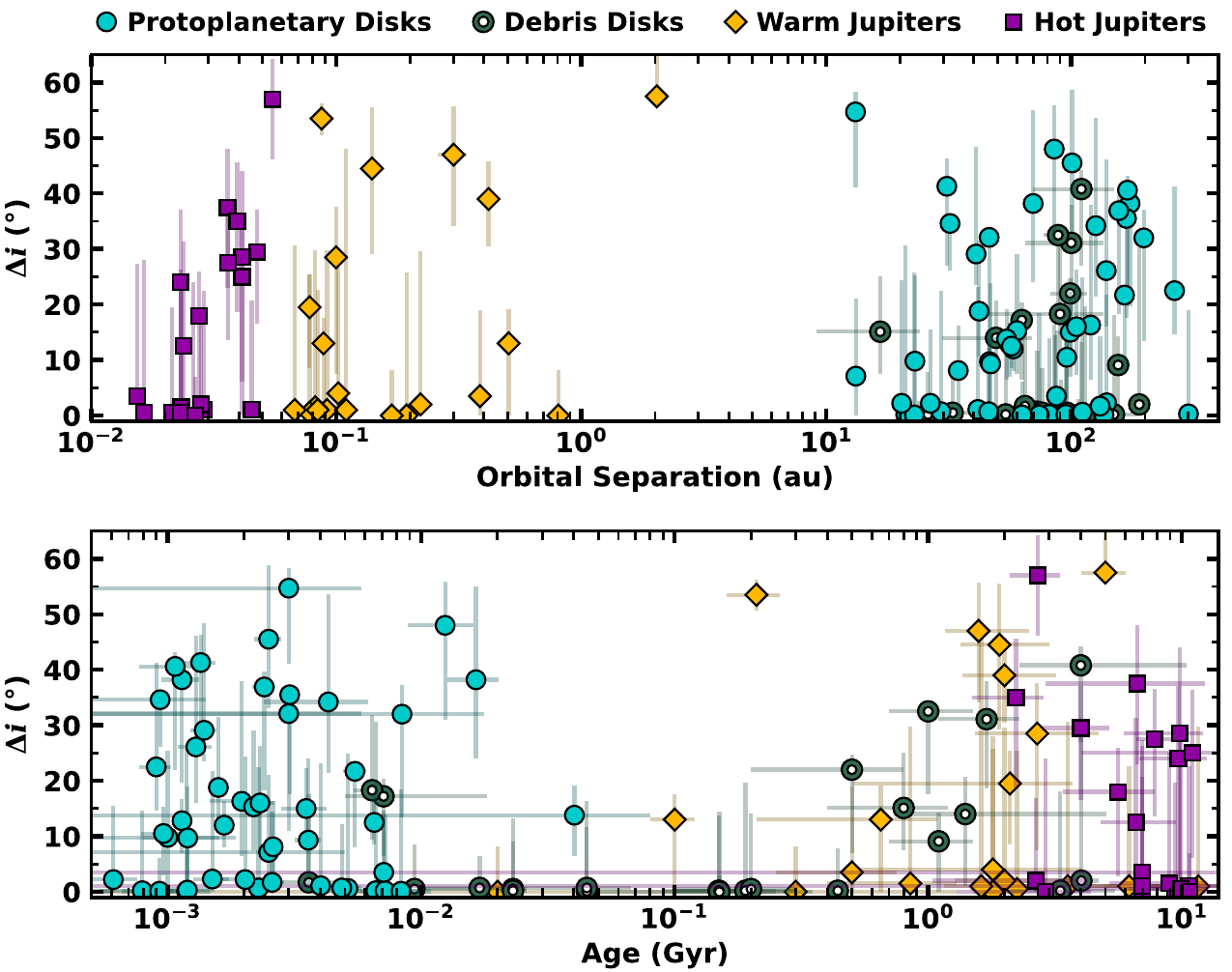}
    \caption{\Deltai as a function of age (upper panel) and orbital separation (lower panel). Points that are solid circles represent the protoplanetary disk sample, points that are rings depict the debris disk sample, and diamond and square points correspond to the warm and hot Jupiter samples, respectively. No obvious trend is apparent across four orders of magnitude in separation and age.}
	\label{fig:age_and_separation_vs_Deltai}
\end{figure*}

\subsection{Comparing Misalignment Rates}
The overall misalignment rates derived from the \Deltai probability distributions of individual systems offer another way to assess whether obliquity evolves over time. If post-formation dynamics result in changes in spin-orbit obliquity, it should be reflected in the misalignment rate, which would increase or decrease depending on the mechanism that works to excite or dampen spin-orbit obliquities. The misalignment rate between samples, however, remains consistent, even when imposing increasingly larger minimum thresholds for counting misaligned systems. The consistency across populations suggests that post-formation dynamical evolution does not significantly impact the overall rate of spin-orbit misalignment in the time between the evolutionary stages represented by the sample populations, at least at the scale of the measurement precision.

\subsection{Single versus Multiple Giant Planets}
In the absence of nearby stellar companions, gravitational interactions between planets are most likely to excite the inclinations of planetary orbits through planet-planet scattering \citep{Chatterjee2008}, ZKL oscillations \citep{vonZeipel1910, Kozai1962, Lidov1962}, and secular chaos \citep{Wu2011} following the dissipation of the protoplanetary disk. Spin-orbit obliquities that diverge from the primordial distribution could have a higher multiplicity rate as a result of excitation mechanisms involving multiple planets. In warm Jupiter systems, 8/\nwj systems have two or more confirmed giant planets (as defined in Section \ref{subsec:J_Sample})\footnote{Based on a query of the NASA Exoplanet Archive \citep{Christiansen2025} as of December 2025.}. In the hot Jupiter sample, one system, WASP-41, is confirmed to host an additional RV-detected giant (WASP-41~c; \citealt{Neveu-VanMalle2016}) with an orbital separation of $\sim$1~au. Based on the transit of WASP-41~b, the minimum obliquity of the system is 29$^{+16}_{-29}$$^{\circ}$. Overall, \Deltai shows no obvious trend in relation to single-planet vs. multi-planet systems. However, this observation does not account for undiscovered outer giant planets, nor whether the orbits of outer giants are inclined with respect to their transiting counterpart. In the absence of a complete picture of outer planets and mutual inclinations, we cannot reliably test whether \Deltai is correlated with planetary multiplicity.

\subsection{Obliquity Evolution}
The consistency in the distribution of \Deltai and misalignment rates across each population suggests there is little or no evolution in stellar obliquity over time, and if large changes do occur, they are rare. Instead, the star-planet obliquity distribution may simply be inherited from the orientation of the protoplanetary disk rather than a result of planet-planet gravitational interactions following the epoch of planet formation. Although individual systems undoubtedly scatter and take on mutual inclinations (e.g., \citealt{Mills2017, BardalezGagliuffi2025}), the magnitude of the change as a population is less than the precision of our observations. We note that the obliquity distributions measured here are a function of \Deltai, which is the lower limit of the true obliquity angle ($\Psi$). The true obliquity distributions are likely greater than shown here, and may possibly deviate significantly between sample populations. However, the consistency in the minimum misalignment rate across each sample as characterized by \Deltai suggests that both the overall rate and degree of misalignment do not increase or decrease between evolutionary stages. While distinct $\Psi$ distributions can project onto similar \Deltai distributions, large-amplitude evolution affecting a substantial fraction of systems would still manifest as an excess of large \Deltai values. The absence of such an excess implies that any population-level evolution must either be modest in amplitude or confined to a minority of systems.

If primordial misalignment sets the distribution of stellar obliquities for Sun-like and low-mass stars, then the observed alignment of many hot-Jupiter systems around cool stars does not necessarily require tidal realignment of the stellar spin axis. Early interpretations attributed the transition near \Teff$\sim$6200~K to efficient tidal realignment in cool stars with substantial convective envelopes \citep{Winn2010, Albrecht2012}. However, tidal realignment of the host star remains theoretically challenging because reorienting the stellar spin without substantial orbital decay requires a large amount of angular momentum transfer from the planetary orbit, resulting in alignment timescales that often exceed 1~Gyr---potentially by several orders of magnitude \citep{Barker2009, Lai2012, Rogers2013}. Because tides tend to remove orbital angular momentum more rapidly than they realign the stellar spin \citep{Barker2009, Rogers2013, Schmidt2026}, it could take longer than the observed system age to align the star from its primordial configuration or erase any further appreciable obliquity excitation (e.g., via planet-planet scattering or ZKL oscillations) that may have occurred following the dissipation of the disk, though the exact timescale will vary depending on the star's assumed tidal quality factor. Moreover, tidal dissipation drives obliquities toward 0$^\circ$, 90$^\circ$, or 180$^\circ$ as opposed to the more moderate obliquities observed here. Therefore, we see no evidence in the hot Jupiter sample in support of a scenario where changing spin-orbit obliquity is driven by tides.

For warm Jupiters, obliquity evolution is unlikely to be driven by tides, with estimated timescales for tidal alignment exceeding hundreds of Gyr \citep{Barker2009, Lai2012, Rogers2013}. In addition, growing observational evidence indicates that warm Jupiters around hot stars tend to remain well aligned \citep{Wang2024, Rusznak2025}, suggesting that the stellar obliquity distribution may instead be inherited from the primordial star–disk configuration rather than produced by subsequent tidal evolution. In this picture, the primary discrepancy lies not with the aligned hot-Jupiter population around cool stars, but rather with the prevalence of highly misaligned hot Jupiters orbiting hot stars.

Obliquity evolution could alternatively occur via secular interactions in multiplanet systems, which can drive oscillations of mutual inclination between an inner and outer planet---and thus, $\Delta i$---on timescales that are shorter than the typical age of the transiting planet sample (e.g., \citealt{BoueFabrycky2014a, BoueFabrycky2014b, BeckerAdams2016}), with amplitudes set by the masses, spacings, and the initial mutual inclination of the planets involved. If secular evolution has occurred, the planets will have been observed at effectively random secular phases, and such evolution could create, erase, or mask differences between populations. However, without complete knowledge of the existence of outer planets in the transiting planet samples, we cannot conclude if the distribution in $\Delta i$ has been altered by secular interactions between planets. Nevertheless, widespread secular excitation to large stellar obliquities could produce an excess of high $\Delta i$ systems relative to the protoplanetary disk population. The absence of such an excess is consistent with secular evolution being modest in amplitude or affecting only a minority of systems, although quantitative constraints require modeling the secular dynamics, viewing geometry, observational selection, and measurement uncertainties. Small-amplitude evolution could therefore remain undetected at our current precision. A related analysis by \citet{Becker2017}, which included WASP-41 from our transiting planet sample, statistically favored low mutual inclinations in a population of hot Jupiter systems with outer companions, providing a complementary population-level argument for relatively coplanar architectures.

The eccentricities of warm Jupiters in this sample provide an additional constraint on their dynamical histories. Their eccentricities span a broad range, reaching values as high as $e$$\approx$0.8, demonstrating that at least some of these systems have undergone substantial dynamical excitation (e.g., \citealt{Dawson2014b, Morgan2025, Morgan2026}). Yet this excitation is not accompanied by a corresponding excess of large minimum obliquities: the (\Deltai) distribution and misalignment rate of warm Jupiters remain broadly consistent with those of the protoplanetary disk population. This combination of high eccentricities and generally low stellar obliquities suggests that the processes responsible for exciting warm Jupiter eccentricities do not necessarily produce comparably large changes in orbital inclination. Dynamical evolution may instead occur preferentially within or near the primordial orbital plane, or inclination excitation may be sufficiently modest or transient that it is not apparent at the precision of the current (\Deltai) measurements. More generally, these results demonstrate that substantial evolution in eccentricity does not require similarly substantial evolution in stellar obliquity.

However, the difference in spatial scales probed by disk observations compared with transiting planets complicates the interpretation of these results, as the flatness of planetary systems has not been determined out to tens or hundreds of au. A comparison of star-disk obliquities to systems with planetary orbits similar to separations resolved in protoplanetary disks observations would offer clearer insights about dynamical evolution as it relates to the outer regions of planetary systems. Increasing the number of directly imaged planets with stellar obliquities will be important to analyze this population. Meanwhile, the separation space between 1--$\sim$10~au will be populated by Gaia DR4 and DR5, which are expected to provide orbital solutions for thousands of giant planets in this range \citep{Perryman2014, Lammers2026}.

% -------------- SUMMARY -------------- %
\section{Summary}\label{sec:Summary}

In this work, we compare minimum obliquity distributions of four distinct planetary system populations––\nppd protoplanetary disk systems, \ndd debris disk systems, \nwj warm Jupiter systems, and \nhj hot Jupiter systems––to test whether stellar obliquity evolves over time. We assess the similarities and differences in the minimum obliquity distributions relative probability distributions and misalignment rates. By both metrics, the minimum obliquity distributions are statistically consistent within 1$\sigma$. The one exception is the $\Delta i$ distribution function for debris disks, which appear somewhat more aligned than the other populations. However, this may be attributed to the smaller uncertainties of individual measurements for that sample.

Several post-formation dynamical mechanisms, including planet–planet scattering and von Zeipel–Kozai-Lidov oscillations, predict the emergence of a larger population of highly misaligned systems over time, particularly among giant planets. Instead, we find that the minimum-obliquity distributions of warm Jupiters and hot Jupiters around cool stars remain broadly consistent with the primordial distribution measured in protoplanetary disks. This suggests that large-scale inclination excitation does not dominate the obliquity evolution of most cool-star giant transiting planet systems, at least at the level detectable with current \Deltai measurements, though relatively moderate evolution ($\lesssim$10$^{\circ}$) remains plausible. An examination of the planetary eccentricities in this work reveals that the eccentricities of warm Jupiters are broadly distributed (up to $e$$\approx$0.8), indicating that these systems have undergone dynamical excitation, but the mechanism for excitation is either coplanar or does not affect orbital inclination in a way that is perceptible in the current dataset. 

The lack of significant differences in minimum obliquity between populations suggests that following the epoch of planet formation, large changes in stellar obliquity may be either uncommon or occur in regions of parameter space not examined in this work. If the obliquity distribution of cool-star systems is largely primordial, then the majority aligned hot-Jupiter population around cool stars may not require large-scale tidal realignment from an initially misaligned configuration. While these findings are consistent with stellar obliquity remaining largely stable over time, the difference in orbital separations between the disk-bearing and planet-bearing populations may result in different processes that affect spin-orbit geometries. Furthermore, transiting hot and warm Jupiters may not trace the orbits of undiscovered outer giant planets, leaving the impact of planet multiplicity and degree of coplanarity open to question. Future observations of wide-separation planets with direct imaging and astrometry will help refine the current mapping of stellar obliquity across space and time.

\section*{Acknowledgments}
L.I.B. is grateful to Joshua N. Winn for helpful discussion about interpretations of \Deltai obtained with the \vsini method and insights about inferences that can be made about their true obliquities. B.P.B. acknowledges support from the National Science Foundation grant AST-1909209, NASA Exoplanet Research Program grant 20-XRP20$\_$2-0119, and the Alfred P. Sloan Foundation. This research has made use of the VizieR catalogue access tool, CDS, Strasbourg, France \citep{10.26093/cds/vizier}. The original description of the VizieR service was published in \citet{vizier2000}. This work made use of the following software: \texttt{Astropy} \citep{Robitaille2021}, \texttt{Numpy} \citep{vanderWalt2011}, \texttt{Scipy} \citep{Virtanen2021}, \texttt{Matplotlib} \citep{Hunter2007}, and \texttt{lightkurve} \citep{LightkurveCollaboration2018}.

\bibliography{references.bib}{}

\begin{thebibliography}{}
\expandafter\ifx\csname natexlab\endcsname\relax\def\natexlab#1{#1}\fi
\providecommand{\url}[1]{\href{#1}{#1}}
\providecommand{\dodoi}[1]{doi:~\href{http://doi.org/#1}{\nolinkurl{#1}}}
\providecommand{\doeprint}[1]{\href{http://ascl.net/#1}{\nolinkurl{http://ascl.net/#1}}}
\providecommand{\doarXiv}[1]{\href{https://arxiv.org/abs/#1}{\nolinkurl{https://arxiv.org/abs/#1}}}

\bibitem[{{Abdurro'uf} {et~al.}(2022){Abdurro'uf}, {Accetta}, {Aerts}, {Silva
  Aguirre}, {Ahumada}, {Ajgaonkar}, {Filiz Ak}, {Alam}, {Allende Prieto},
  {Almeida}, {Anders}, {Anderson}, {Andrews}, {Anguiano}, {Aquino-Ort{\'\i}z},
  {Arag{\'o}n-Salamanca}, {Argudo-Fern{\'a}ndez}, {Ata}, {Aubert},
  {Avila-Reese}, {Badenes}, {Barb{\'a}}, {Barger}, {Barrera-Ballesteros},
  {Beaton}, {Beers}, {Belfiore}, {Bender}, {Bernardi}, {Bershady}, {Beutler},
  {Bidin}, {Bird}, {Bizyaev}, {Blanc}, {Blanton}, {Boardman}, {Bolton},
  {Boquien}, {Borissova}, {Bovy}, {Brandt}, {Brown}, {Brownstein}, {Brusa},
  {Buchner}, {Bundy}, {Burchett}, {Bureau}, {Burgasser}, {Cabang}, {Campbell},
  {Cappellari}, {Carlberg}, {Wanderley}, {Carrera}, {Cash}, {Chen}, {Chen},
  {Cherinka}, {Chiappini}, {Choi}, {Chojnowski}, {Chung}, {Clerc}, {Cohen},
  {Comerford}, {Comparat}, {da Costa}, {Covey}, {Crane}, {Cruz-Gonzalez},
  {Culhane}, {Cunha}, {Dai}, {Damke}, {Darling}, {Davidson}, {Davies},
  {Dawson}, {De Lee}, {Diamond-Stanic}, {Cano-D{\'\i}az}, {S{\'a}nchez},
  {Donor}, {Duckworth}, {Dwelly}, {Eisenstein}, {Elsworth}, {Emsellem},
  {Eracleous}, {Escoffier}, {Fan}, {Farr}, {Feng}, {Fern{\'a}ndez-Trincado},
  {Feuillet}, {Filipp}, {Fillingham}, {Frinchaboy}, {Fromenteau}, {Galbany},
  {Garc{\'\i}a}, {Garc{\'\i}a-Hern{\'a}ndez}, {Ge}, {Geisler}, {Gelfand},
  {G{\'e}ron}, {Gibson}, {Goddy}, {Godoy-Rivera}, {Grabowski}, {Green},
  {Greener}, {Grier}, {Griffith}, {Guo}, {Guy}, {Hadjara}, {Harding},
  {Hasselquist}, {Hayes}, {Hearty}, {Hern{\'a}ndez}, {Hill}, {Hogg},
  {Holtzman}, {Horta}, {Hsieh}, {Hsu}, {Hsu}, {Huber}, {Huertas-Company},
  {Hutchinson}, {Hwang}, {Ibarra-Medel}, {Chitham}, {Ilha}, {Imig}, {Jaekle},
  {Jayasinghe}, {Ji}, {Johnson}, {Jones}, {J{\"o}nsson}, {Katkov}, {Khalatyan},
  {Kinemuchi}, {Kisku}, {Knapen}, {Kneib}, {Kollmeier}, {Kong}, {Kounkel},
  {Kreckel}, {Krishnarao}, {Lacerna}, {Lane}, {Langgin}, {Lavender}, {Law},
  {Lazarz}, {Leung}, {Leung}, {Lewis}, {Li}, {Li}, {Lian}, {Liang}, {Lin},
  {Lin}, {Lin}, {Lintott}, {Long}, {Longa-Pe{\~n}a}, {L{\'o}pez-Cob{\'a}},
  {Lu}, {Lundgren}, {Luo}, {Mackereth}, {de la Macorra}, {Mahadevan},
  {Majewski}, {Manchado}, {Mandeville}, {Maraston}, {Margalef-Bentabol},
  {Masseron}, {Masters}, {Mathur}, {McDermid}, {Mckay}, {Merloni},
  {Merrifield}, {Meszaros}, {Miglio}, {Di Mille}, {Minniti}, {Minsley}, \&
  {Monachesi}}]{Abdurro'uf2022}
{Abdurro'uf}, {Accetta}, K., {Aerts}, C., {et~al.} 2022, The Astrophysical
  Journal Supplement Series, 259, 35, \dodoi{10.3847/1538-4365/ac4414}

\bibitem[{{Aerts}(2021)}]{Aerts2021}
{Aerts}, C. 2021, Reviews of Modern Physics, 93, 015001,
  \dodoi{10.1103/RevModPhys.93.015001}

\bibitem[{{Aguilera-G{\'o}mez} {et~al.}(2018){Aguilera-G{\'o}mez},
  {Ram{\'\i}rez}, \& {Chanam{\'e}}}]{Aguilera-Gomez2018}
{Aguilera-G{\'o}mez}, C., {Ram{\'\i}rez}, I., \& {Chanam{\'e}}, J. 2018,
  Astronomy and Astrophysics, 614, A55, \dodoi{10.1051/0004-6361/201732209}

\bibitem[{{Albrecht} {et~al.}(2007){Albrecht}, {Reffert}, {Snellen},
  {Quirrenbach}, \& {Mitchell}}]{Albrecht2007}
{Albrecht}, S., {Reffert}, S., {Snellen}, I., {Quirrenbach}, A., \& {Mitchell},
  D.~S. 2007, Astronomy and Astrophysics, 474, 565,
  \dodoi{10.1051/0004-6361:20077953}

\bibitem[{{Albrecht} {et~al.}(2012){Albrecht}, {Winn}, {Johnson}, {Howard},
  {Marcy}, {Butler}, {Arriagada}, {Crane}, {Shectman}, {Thompson}, {Hirano},
  {Bakos}, \& {Hartman}}]{Albrecht2012}
{Albrecht}, S., {Winn}, J.~N., {Johnson}, J.~A., {et~al.} 2012, \apj, 757, 18,
  \dodoi{10.1088/0004-637X/757/1/18}

\bibitem[{{Albrecht} {et~al.}(2022){Albrecht}, {Dawson}, \&
  {Winn}}]{Albrecht2022}
{Albrecht}, S.~H., {Dawson}, R.~I., \& {Winn}, J.~N. 2022, \pasp, 134, 082001,
  \dodoi{10.1088/1538-3873/ac6c09}

\bibitem[{{Andretta} {et~al.}(2017){Andretta}, {Giampapa}, {Covino}, {Reiners},
  \& {Beeck}}]{Andretta2017}
{Andretta}, V., {Giampapa}, M.~S., {Covino}, E., {Reiners}, A., \& {Beeck}, B.
  2017, The Astrophysical Journal, 839, 97, \dodoi{10.3847/1538-4357/aa6a14}

\bibitem[{{Baliunas} {et~al.}(1996){Baliunas}, {Sokoloff}, \&
  {Soon}}]{Baliunas1996}
{Baliunas}, S., {Sokoloff}, D., \& {Soon}, W. 1996, The Astrophysical Journal,
  457, L99, \dodoi{10.1086/309891}

\bibitem[{{Balmer} {et~al.}(2026){Balmer}, {Pueyo}, {Messier}, {Bruinsma},
  {Jones}, {Matuszewska}, {Perrin}, {Girard}, {Leisenring}, {Lawson}, {van der
  Marel}, {Kammerer}, {Carter}, {M{\^a}lin}, {Ward-Duong}, {Hoch}, {Rickman},
  \& {Seager}}]{Balmer2026}
{Balmer}, W.~O., {Pueyo}, L., {Messier}, A., {et~al.} 2026, The Astrophysical
  Journal, 1001, L26, \dodoi{10.3847/2041-8213/ae374a}

\bibitem[{{Bardalez Gagliuffi} {et~al.}(2025){Bardalez Gagliuffi}, {Balmer},
  {Pueyo}, {Brandt}, {Giovinazzi}, {Millholland}, {Black}, {Lu}, {Rice},
  {Mang}, {Morley}, {Lacy}, {Girard}, {Matthews}, {Carter}, {Bowler},
  {Faherty}, {Fontanive}, \& {Rickman}}]{BardalezGagliuffi2025}
{Bardalez Gagliuffi}, D.~C., {Balmer}, W.~O., {Pueyo}, L., {et~al.} 2025, The
  Astrophysical Journal, 988, L18, \dodoi{10.3847/2041-8213/ade30f}

\bibitem[{{Barker} \& {Ogilvie}(2009)}]{Barker2009}
{Barker}, A.~J., \& {Ogilvie}, G.~I. 2009, Monthly Notices of the Royal
  Astronomical Society, 395, 2268, \dodoi{10.1111/j.1365-2966.2009.14694.x}

\bibitem[{{Becker} \& {Adams}(2016)}]{BeckerAdams2016}
{Becker}, J.~C., \& {Adams}, F.~C. 2016, \mnras, 455, 2980,
  \dodoi{10.1093/mnras/stv2444}

\bibitem[{{Becker} {et~al.}(2017){Becker}, {Vanderburg}, {Adams}, {Khain}, \&
  {Bryan}}]{Becker2017}
{Becker}, J.~C., {Vanderburg}, A., {Adams}, F.~C., {Khain}, T., \& {Bryan}, M.
  2017, \aj, 154, 230, \dodoi{10.3847/1538-3881/aa9176}

\bibitem[{{Bell} {et~al.}(2015){Bell}, {Mamajek}, \& {Naylor}}]{Bell2015}
{Bell}, C. P.~M., {Mamajek}, E.~E., \& {Naylor}, T. 2015, Monthly Notices of
  the Royal Astronomical Society, 454, 593, \dodoi{10.1093/mnras/stv1981}

\bibitem[{{Beuzit} {et~al.}(2008){Beuzit}, {Feldt}, {Dohlen}, {Mouillet},
  {Puget}, {Wildi}, {Abe}, {Antichi}, {Baruffolo}, {Baudoz}, {Boccaletti},
  {Carbillet}, {Charton}, {Claudi}, {Downing}, {Fabron}, {Feautrier},
  {Fedrigo}, {Fusco}, {Gach}, {Gratton}, {Henning}, {Hubin}, {Joos}, {Kasper},
  {Langlois}, {Lenzen}, {Moutou}, {Pavlov}, {Petit}, {Pragt}, {Rabou}, {Rigal},
  {Roelfsema}, {Rousset}, {Saisse}, {Schmid}, {Stadler}, {Thalmann}, {Turatto},
  {Udry}, {Vakili}, \& {Waters}}]{Beuzit2008}
{Beuzit}, J.-L., {Feldt}, M., {Dohlen}, K., {et~al.} 2008, in Society of
  Photo-Optical Instrumentation Engineers (SPIE) Conference Series, Vol. 7014,
  Ground-based and Airborne Instrumentation for Astronomy II, ed. I.~S.
  {McLean} \& M.~M. {Casali}, 701418, \dodoi{10.1117/12.790120}

\bibitem[{{Biddle} {et~al.}(2025){Biddle}, {Bowler}, {Morgan}, {Tran}, \&
  {Wu}}]{Biddle2025}
{Biddle}, L.~I., {Bowler}, B.~P., {Morgan}, M., {Tran}, Q.~H., \& {Wu}, Y.-L.
  2025, Nature, 644, 356, \dodoi{10.1038/s41586-025-09324-0}

\bibitem[{{Boccaletti} {et~al.}(2019){Boccaletti}, {Th{\'e}bault}, {Pawellek},
  {Lagrange}, {Galicher}, {Desidera}, {Milli}, {Kral}, {Bonnefoy}, {Augereau},
  {Maire}, {Henning}, {Beust}, {Rodet}, {Avenhaus}, {Bhowmik}, {Bonavita},
  {Chauvin}, {Cheetham}, {Cudel}, {Feldt}, {Gratton}, {Hagelberg},
  {Janin-Potiron}, {Langlois}, {Menard}, {Mesa}, {Meyer}, {Peretti}, {Perrot},
  {Schmidt}, {Sissa}, {Vigan}, {Rickman}, {Magnard}, {Maurel},
  {Moeller-Nilsson}, {Perret}, \& {Sauvage}}]{Boccaletti2019}
{Boccaletti}, A., {Th{\'e}bault}, P., {Pawellek}, N., {et~al.} 2019, Astronomy
  and Astrophysics, 625, A21, \dodoi{10.1051/0004-6361/201935135}

\bibitem[{{Booth} {et~al.}(2023){Booth}, {Pearce}, {Krivov}, {Wyatt}, {Dent},
  {Hales}, {Lestrade}, {Cruz-S{\'a}enz de Miera}, {Faramaz}, {L{\"o}hne}, \&
  {Chavez-Dagostino}}]{Booth2023}
{Booth}, M., {Pearce}, T.~D., {Krivov}, A.~V., {et~al.} 2023, Monthly Notices
  of the Royal Astronomical Society, 521, 6180, \dodoi{10.1093/mnras/stad938}

\bibitem[{{Borisov} {et~al.}(2023){Borisov}, {Chilingarian}, {Rubtsov},
  {Ledoux}, {Melo}, {Grishin}, {Katkov}, {Goradzhanov}, {Afanasiev},
  {Kasparova}, \& {Saburova}}]{Borisov2023}
{Borisov}, S.~B., {Chilingarian}, I.~V., {Rubtsov}, E.~V., {et~al.} 2023, The
  Astrophysical Journal Supplement Series, 266, 11,
  \dodoi{10.3847/1538-4365/acc321}

\bibitem[{{Bou{\'e}} \& {Fabrycky}(2014{\natexlab{a}})}]{BoueFabrycky2014a}
{Bou{\'e}}, G., \& {Fabrycky}, D.~C. 2014{\natexlab{a}}, \apj, 789, 110,
  \dodoi{10.1088/0004-637X/789/2/110}

\bibitem[{{Bou{\'e}} \& {Fabrycky}(2014{\natexlab{b}})}]{BoueFabrycky2014b}
---. 2014{\natexlab{b}}, \apj, 789, 111, \dodoi{10.1088/0004-637X/789/2/111}

\bibitem[{{Bowler} {et~al.}(2023){Bowler}, {Tran}, {Zhang}, {Morgan}, {Ashok},
  {Blunt}, {Bryan}, {Evans}, {Franson}, {Huber}, {Nagpal}, {Wu}, \&
  {Zhou}}]{Bowler2023}
{Bowler}, B.~P., {Tran}, Q.~H., {Zhang}, Z., {et~al.} 2023, \aj, 165, 164,
  \dodoi{10.3847/1538-3881/acbd34}

\bibitem[{{Brewer} \& {Donovan}(2015)}]{Brewer2015}
{Brewer}, B.~J., \& {Donovan}, C.~P. 2015, Monthly Notices of the Royal
  Astronomical Society, 448, 3206, \dodoi{10.1093/mnras/stv199}

\bibitem[{{Brewer} {et~al.}(2016){Brewer}, {Fischer}, {Valenti}, \&
  {Piskunov}}]{Brewer2016}
{Brewer}, J.~M., {Fischer}, D.~A., {Valenti}, J.~A., \& {Piskunov}, N. 2016,
  \apjs, 225, 32, \dodoi{10.3847/0067-0049/225/2/32}

\bibitem[{{Buder} {et~al.}(2019){Buder}, {Lind}, {Ness}, {Asplund}, {Duong},
  {Lin}, {Kos}, {Casagrande}, {Casey}, {Bland-Hawthorn}, {de Silva}, {D'Orazi},
  {Freeman}, {Martell}, {Schlesinger}, {Sharma}, {Simpson}, {Zucker},
  {Zwitter}, {{\v{C}}otar}, {Dotter}, {Hayden}, {Hyde}, {Kafle}, {Lewis},
  {Nataf}, {Nordlander}, {Reid}, {Rix}, {Sk{\'u}lad{\'o}ttir}, {Stello},
  {Ting}, {Traven}, {Wyse}, \& {GALAH Collaboration}}]{Buder2019}
{Buder}, S., {Lind}, K., {Ness}, M.~K., {et~al.} 2019, Astronomy and
  Astrophysics, 624, A19, \dodoi{10.1051/0004-6361/201833218}

\bibitem[{{Casagrande} {et~al.}(2011){Casagrande}, {Sch{\"o}nrich}, {Asplund},
  {Cassisi}, {Ram{\'\i}rez}, {Mel{\'e}ndez}, {Bensby}, \&
  {Feltzing}}]{Casagrande2011}
{Casagrande}, L., {Sch{\"o}nrich}, R., {Asplund}, M., {et~al.} 2011, Astronomy
  and Astrophysics, 530, A138, \dodoi{10.1051/0004-6361/201016276}

\bibitem[{{Chatterjee} {et~al.}(2008){Chatterjee}, {Ford}, {Matsumura}, \&
  {Rasio}}]{Chatterjee2008}
{Chatterjee}, S., {Ford}, E.~B., {Matsumura}, S., \& {Rasio}, F.~A. 2008, \apj,
  686, 580, \dodoi{10.1086/590227}

\bibitem[{{Chatterjee} {et~al.}(2011){Chatterjee}, {Ford}, \&
  {Rasio}}]{Chatterjee2011}
{Chatterjee}, S., {Ford}, E.~B., \& {Rasio}, F.~A. 2011, in IAU Symposium, Vol.
  276, The Astrophysics of Planetary Systems: Formation, Structure, and
  Dynamical Evolution, ed. A.~{Sozzetti}, M.~G. {Lattanzi}, \& A.~P. {Boss},
  225--229, \dodoi{10.1017/S1743921311020229}

\bibitem[{{Chen} {et~al.}(2011){Chen}, {Mamajek}, {Bitner}, {Pecaut}, {Su}, \&
  {Weinberger}}]{Chen2011}
{Chen}, C.~H., {Mamajek}, E.~E., {Bitner}, M.~A., {et~al.} 2011, The
  Astrophysical Journal, 738, 122, \dodoi{10.1088/0004-637X/738/2/122}

\bibitem[{{Choquet} {et~al.}(2016){Choquet}, {Perrin}, {Chen}, {Soummer},
  {Pueyo}, {Hagan}, {Gofas-Salas}, {Rajan}, {Golimowski}, {Hines}, {Schneider},
  {Mazoyer}, {Augereau}, {Debes}, {Stark}, {Wolff}, {N'Diaye}, \&
  {Hsiao}}]{Choquet2016}
{Choquet}, {\'E}., {Perrin}, M.~D., {Chen}, C.~H., {et~al.} 2016, The
  Astrophysical Journal, 817, L2, \dodoi{10.3847/2041-8205/817/1/L2}

\bibitem[{{Choquet} {et~al.}(2018){Choquet}, {Bryden}, {Perrin}, {Soummer},
  {Augereau}, {Chen}, {Debes}, {Gofas-Salas}, {Hagan}, {Hines}, {Mawet},
  {Morales}, {Pueyo}, {Rajan}, {Ren}, {Schneider}, {Stark}, \&
  {Wolff}}]{Choquet2018}
{Choquet}, {\'E}., {Bryden}, G., {Perrin}, M.~D., {et~al.} 2018, The
  Astrophysical Journal, 854, 53, \dodoi{10.3847/1538-4357/aaa892}

\bibitem[{{Christiansen} {et~al.}(2025){Christiansen}, {McElroy}, {Harbut},
  {Ciardi}, {Crane}, {Good}, {Hardegree-Ullman}, {Kesseli}, {Lund}, {Lynn},
  {Muthiar}, {Nilsson}, {Oluyide}, {Papin}, {Rivera}, {Swain}, {Susemiehl},
  {Tam}, {van Eyken}, \& {Beichman}}]{Christiansen2025}
{Christiansen}, J.~L., {McElroy}, D.~L., {Harbut}, M., {et~al.} 2025, The
  Planetary Science Journal, 6, 186, \dodoi{10.3847/PSJ/ade3c2}

\bibitem[{{Collier Cameron} {et~al.}(2010){Collier Cameron}, {Bruce}, {Miller},
  {Triaud}, \& {Queloz}}]{CollierCameron2010}
{Collier Cameron}, A., {Bruce}, V.~A., {Miller}, G.~R.~M., {Triaud},
  A.~H.~M.~J., \& {Queloz}, D. 2010, Monthly Notices of the Royal Astronomical
  Society, 403, 151, \dodoi{10.1111/j.1365-2966.2009.16131.x}

\bibitem[{{Costa Silva} {et~al.}(2020){Costa Silva}, {Delgado Mena}, \&
  {Tsantaki}}]{CostaSilva2020}
{Costa Silva}, A.~R., {Delgado Mena}, E., \& {Tsantaki}, M. 2020, Astronomy and
  Astrophysics, 634, A136, \dodoi{10.1051/0004-6361/201936523}

\bibitem[{{Cronin-Coltsmann} {et~al.}(2021){Cronin-Coltsmann}, {Kennedy},
  {Kalas}, {Milli}, {Clarke}, {Duch{\^e}ne}, {Greaves}, {Lawler}, {Lestrade},
  {Matthews}, {Shannon}, \& {Wyatt}}]{Cronin-Coltsmann2021}
{Cronin-Coltsmann}, P.~F., {Kennedy}, G.~M., {Kalas}, P., {et~al.} 2021,
  Monthly Notices of the Royal Astronomical Society, 504, 4497,
  \dodoi{10.1093/mnras/stab1237}

\bibitem[{{Crotts} {et~al.}(2024){Crotts}, {Matthews}, {Duch{\^e}ne},
  {Esposito}, {Dong}, {Hom}, {Oppenheimer}, {Rice}, {Wolff}, {Chen}, {Do
  {\'O}}, {Kalas}, {Lewis}, {Weinberger}, {Wilner}, {Ammons}, {Arriaga}, {De
  Rosa}, {Debes}, {Fitzgerald}, {Gonzales}, {Hines}, {Hinkley}, {Hughes},
  {Kolokolova}, {Lee}, {L{\'o}pez}, {Macintosh}, {Mazoyer}, {Metchev},
  {Millar-Blanchaer}, {Nielsen}, {Patience}, {Perrin}, {Pueyo},
  {Rantakyr{\"o}}, {Ren}, {Schneider}, {Soummer}, \& {Stark}}]{Crotts2024}
{Crotts}, K.~A., {Matthews}, B.~C., {Duch{\^e}ne}, G., {et~al.} 2024, The
  Astrophysical Journal, 961, 245, \dodoi{10.3847/1538-4357/ad0e69}

\bibitem[{{Cutispoto} {et~al.}(2002){Cutispoto}, {Pastori}, {Pasquini}, {de
  Medeiros}, {Tagliaferri}, \& {Andersen}}]{Cutispoto2002}
{Cutispoto}, G., {Pastori}, L., {Pasquini}, L., {et~al.} 2002, Astronomy and
  Astrophysics, 384, 491, \dodoi{10.1051/0004-6361:20020040}

\bibitem[{{da Silva} {et~al.}(2009){da Silva}, {Torres}, {de La Reza}, {Quast},
  {Melo}, \& {Sterzik}}]{daSilva2009}
{da Silva}, L., {Torres}, C.~A.~O., {de La Reza}, R., {et~al.} 2009, Astronomy
  and Astrophysics, 508, 833, \dodoi{10.1051/0004-6361/200911736}

\bibitem[{{Dawson}(2014)}]{Dawson2014a}
{Dawson}, R.~I. 2014, \apjl, 790, L31, \dodoi{10.1088/2041-8205/790/2/L31}

\bibitem[{{Dawson} \& {Johnson}(2018)}]{Dawson2018}
{Dawson}, R.~I., \& {Johnson}, J.~A. 2018, Annual Review of Astronomy and
  Astrophysics, 56, 175, \dodoi{10.1146/annurev-astro-081817-051853}

\bibitem[{{Dawson} {et~al.}(2014){Dawson}, {Johnson}, {Fabrycky},
  {Foreman-Mackey}, {Murray-Clay}, {Buchhave}, {Cargile}, {Clubb}, {Fulton},
  {Hebb}, {Howard}, {Huber}, {Shporer}, \& {Valenti}}]{Dawson2014b}
{Dawson}, R.~I., {Johnson}, J.~A., {Fabrycky}, D.~C., {et~al.} 2014, The
  Astrophysical Journal, 791, 89, \dodoi{10.1088/0004-637X/791/2/89}

\bibitem[{{dos Santos} {et~al.}(2016){dos Santos}, {Mel{\'e}ndez}, {do
  Nascimento}, {Bedell}, {Ram{\'\i}rez}, {Bean}, {Asplund}, {Spina},
  {Dreizler}, {Alves-Brito}, \& {Casagrande}}]{dosSantos2016}
{dos Santos}, L.~A., {Mel{\'e}ndez}, J., {do Nascimento}, J.-D., {et~al.} 2016,
  Astronomy and Astrophysics, 592, A156, \dodoi{10.1051/0004-6361/201628558}

\bibitem[{{Eiroa} {et~al.}(2013){Eiroa}, {Marshall}, {Mora}, {Montesinos},
  {Absil}, {Augereau}, {Bayo}, {Bryden}, {Danchi}, {del Burgo}, {Ertel},
  {Fridlund}, {Heras}, {Krivov}, {Launhardt}, {Liseau}, {L{\"o}hne},
  {Maldonado}, {Pilbratt}, {Roberge}, {Rodmann}, {Sanz-Forcada}, {Solano},
  {Stapelfeldt}, {Th{\'e}bault}, {Wolf}, {Ardila}, {Ar{\'e}valo}, {Beichmann},
  {Faramaz}, {Gonz{\'a}lez-Garc{\'\i}a}, {Guti{\'e}rrez}, {Lebreton},
  {Mart{\'\i}nez-Arn{\'a}iz}, {Meeus}, {Montes}, {Olofsson}, {Su}, {White},
  {Barrado}, {Fukagawa}, {Gr{\"u}n}, {Kamp}, {Lorente}, {Morbidelli},
  {M{\"u}ller}, {Mutschke}, {Nakagawa}, {Ribas}, \& {Walker}}]{Eiroa2013}
{Eiroa}, C., {Marshall}, J.~P., {Mora}, A., {et~al.} 2013, Astronomy and
  Astrophysics, 555, A11, \dodoi{10.1051/0004-6361/201321050}

\bibitem[{{Esplin} \& {Luhman}(2020)}]{Esplin2020}
{Esplin}, T.~L., \& {Luhman}, K.~L. 2020, Astronomical Journal, 159, 282,
  \dodoi{10.3847/1538-3881/ab8dbd}

\bibitem[{{Fabrycky} \& {Tremaine}(2007)}]{Fabrycky2007}
{Fabrycky}, D., \& {Tremaine}, S. 2007, \apj, 669, 1298, \dodoi{10.1086/521702}

\bibitem[{{Faramaz} {et~al.}(2019){Faramaz}, {Krist}, {Stapelfeldt}, {Bryden},
  {Mamajek}, {Matr{\`a}}, {Booth}, {Flaherty}, {Hales}, {Hughes}, {Bayo},
  {Casassus}, {Cuadra}, {Olofsson}, {Su}, \& {Wilner}}]{Faramaz2019}
{Faramaz}, V., {Krist}, J., {Stapelfeldt}, K.~R., {et~al.} 2019, The
  Astronomical Journal, 158, 162, \dodoi{10.3847/1538-3881/ab3ec1}

\bibitem[{{Fischer} \& {Valenti}(2005)}]{Fischer2005}
{Fischer}, D.~A., \& {Valenti}, J. 2005, The Astrophysical Journal, 622, 1102,
  \dodoi{10.1086/428383}

\bibitem[{{Frey} {et~al.}(1991){Frey}, {Grim}, {Hall}, {Mattingly}, {Robb},
  {Wood}, \& {Zeigler}}]{Frey1991}
{Frey}, G.~J., {Grim}, B., {Hall}, D.~S., {et~al.} 1991, The Astronomical
  Journal, 102, 1813, \dodoi{10.1086/116005}

\bibitem[{{Fuhrmann} {et~al.}(2017){Fuhrmann}, {Chini}, {Kaderhandt}, \&
  {Chen}}]{Fuhrmann2017}
{Fuhrmann}, K., {Chini}, R., {Kaderhandt}, L., \& {Chen}, Z. 2017, The
  Astrophysical Journal, 836, 139, \dodoi{10.3847/1538-4357/836/1/139}

\bibitem[{{Fusco} {et~al.}(2014){Fusco}, {Sauvage}, {Petit}, {Costille},
  {Dohlen}, {Mouillet}, {Beuzit}, {Kasper}, {Suarez}, {Soenke}, {Fedrigo},
  {Downing}, {Baudoz}, {Sevin}, {Perret}, {Barufolo}, {Salasnich}, {Puget},
  {Feautrier}, {Rochat}, {Moulin}, {Deboulb{\'e}}, {Hugot}, {Vigan}, {Mawet},
  {Girard}, \& {Hubin}}]{Fusco2014}
{Fusco}, T., {Sauvage}, J.~F., {Petit}, C., {et~al.} 2014, in Society of
  Photo-Optical Instrumentation Engineers (SPIE) Conference Series, Vol. 9148,
  Adaptive Optics Systems IV, ed. E.~{Marchetti}, L.~M. {Close}, \& J.-P.
  {Vran}, 91481U, \dodoi{10.1117/12.2055423}

\bibitem[{{Galli} {et~al.}(2020){Galli}, {Bouy}, {Olivares}, {Miret-Roig},
  {Vieira}, {Sarro}, {Barrado}, {Berihuete}, {Bertout}, {Bertin}, \&
  {Cuillandre}}]{Galli2020}
{Galli}, P.~A.~B., {Bouy}, H., {Olivares}, J., {et~al.} 2020, \aap, 643, A148,
  \dodoi{10.1051/0004-6361/202038717}

\bibitem[{{Galli} {et~al.}(2021){Galli}, {Bouy}, {Olivares}, {Miret-Roig},
  {Sarro}, {Barrado}, {Berihuete}, {Bertin}, \& {Cuillandre}}]{Galli2021}
---. 2021, \aap, 646, A46, \dodoi{10.1051/0004-6361/202039395}

\bibitem[{{Giguere} {et~al.}(2016){Giguere}, {Fischer}, {Zhang}, {Matthews},
  {Cameron}, \& {Henry}}]{Giguere2016}
{Giguere}, M.~J., {Fischer}, D.~A., {Zhang}, C. X.~Y., {et~al.} 2016, The
  Astrophysical Journal, 824, 150, \dodoi{10.3847/0004-637X/824/2/150}

\bibitem[{{Gonzalez} {et~al.}(2010){Gonzalez}, {Carlson}, \&
  {Tobin}}]{Gonzalez2010}
{Gonzalez}, G., {Carlson}, M.~K., \& {Tobin}, R.~W. 2010, Monthly Notices of
  the Royal Astronomical Society, 403, 1368,
  \dodoi{10.1111/j.1365-2966.2009.16195.x}

\bibitem[{{Grandjean} {et~al.}(2021){Grandjean}, {Lagrange}, {Meunier},
  {Rubini}, {Desidera}, {Galland}, {Borgniet}, {Zicher}, {Messina}, {Chauvin},
  {Sterzik}, \& {Pantoja}}]{Grandjean2021}
{Grandjean}, A., {Lagrange}, A.~M., {Meunier}, N., {et~al.} 2021, Astronomy and
  Astrophysics, 650, A39, \dodoi{10.1051/0004-6361/202039672}

\bibitem[{{Grandjean} {et~al.}(2023){Grandjean}, {Lagrange}, {Meunier},
  {Chauvin}, {Borgniet}, {Desidera}, {Galland}, {Kiefer}, {Messina},
  {Iglesias}, {Nicholson}, {Pantoja}, {Rubini}, {Sedaghati}, {Sterzik}, \&
  {Zicher}}]{Grandjean2023}
---. 2023, Astronomy and Astrophysics, 669, A12,
  \dodoi{10.1051/0004-6361/202141235}

\bibitem[{{Guillout} {et~al.}(2009){Guillout}, {Klutsch}, {Frasca}, {Freire
  Ferrero}, {Marilli}, {Mignemi}, {Biazzo}, {Bouvier}, {Monier}, {Motch}, \&
  {Sterzik}}]{Guillout2009}
{Guillout}, P., {Klutsch}, A., {Frasca}, A., {et~al.} 2009, Astronomy and
  Astrophysics, 504, 829, \dodoi{10.1051/0004-6361/200811313}

\bibitem[{{Hale}(1994)}]{Hale1994}
{Hale}, A. 1994, The Astronomical Journal, 107, 306, \dodoi{10.1086/116855}

\bibitem[{{Herrero} {et~al.}(2012){Herrero}, {Ribas}, {Jordi}, {Guinan}, \&
  {Engle}}]{Herrero2012}
{Herrero}, E., {Ribas}, I., {Jordi}, C., {Guinan}, E.~F., \& {Engle}, S.~G.
  2012, Astronomy and Astrophysics, 537, A147,
  \dodoi{10.1051/0004-6361/201117809}

\bibitem[{{Hillenbrand} {et~al.}(2008){Hillenbrand}, {Carpenter}, {Kim},
  {Meyer}, {Backman}, {Moro-Mart{\'\i}n}, {Hollenbach}, {Hines}, {Pascucci}, \&
  {Bouwman}}]{Hillenbrand2008}
{Hillenbrand}, L.~A., {Carpenter}, J.~M., {Kim}, J.~S., {et~al.} 2008, The
  Astrophysical Journal, 677, 630, \dodoi{10.1086/529027}

\bibitem[{{Hojjatpanah} {et~al.}(2019){Hojjatpanah}, {Figueira}, {Santos},
  {Adibekyan}, {Sousa}, {Delgado-Mena}, {Alibert}, {Cristiani}, {Gonz{\'a}lez
  Hern{\'a}ndez}, {Lanza}, {Di Marcantonio}, {Martins}, {Micela}, {Molaro},
  {Neves}, {Oshagh}, {Pepe}, {Poretti}, {Rojas-Ayala}, {Rebolo}, {Su{\'a}rez
  Mascare{\~n}o}, \& {Zapatero Osorio}}]{Hojjatpanah2019}
{Hojjatpanah}, S., {Figueira}, P., {Santos}, N.~C., {et~al.} 2019, Astronomy
  and Astrophysics, 629, A80, \dodoi{10.1051/0004-6361/201834729}

\bibitem[{{Hojjatpanah} {et~al.}(2020){Hojjatpanah}, {Oshagh}, {Figueira},
  {Santos}, {Amazo-G{\'o}mez}, {Sousa}, {Adibekyan}, {Akinsanmi}, {Demangeon},
  {Faria}, {Gomes da Silva}, \& {Meunier}}]{Hojjatpanah2020}
{Hojjatpanah}, S., {Oshagh}, M., {Figueira}, P., {et~al.} 2020, Astronomy and
  Astrophysics, 639, A35, \dodoi{10.1051/0004-6361/202038035}

\bibitem[{{Holman} \& {Wiegert}(1999)}]{Holman1999}
{Holman}, M.~J., \& {Wiegert}, P.~A. 1999, The Astronomical Journal, 117, 621,
  \dodoi{10.1086/300695}

\bibitem[{{Houdebine} {et~al.}(2017){Houdebine}, {Mullan}, {Bercu}, {Paletou},
  \& {Gebran}}]{Houdebine2017}
{Houdebine}, E.~R., {Mullan}, D.~J., {Bercu}, B., {Paletou}, F., \& {Gebran},
  M. 2017, The Astrophysical Journal, 837, 96, \dodoi{10.3847/1538-4357/aa5cad}

\bibitem[{{Houdebine} {et~al.}(2016){Houdebine}, {Mullan}, {Paletou}, \&
  {Gebran}}]{Houdebine2016}
{Houdebine}, E.~R., {Mullan}, D.~J., {Paletou}, F., \& {Gebran}, M. 2016, The
  Astrophysical Journal, 822, 97, \dodoi{10.3847/0004-637X/822/2/97}

\bibitem[{{Huber} {et~al.}(2017){Huber}, {Zinn}, {Bojsen-Hansen},
  {Pinsonneault}, {Sahlholdt}, {Serenelli}, {Silva Aguirre}, {Stassun},
  {Stello}, {Tayar}, {Bastien}, {Bedding}, {Buchhave}, {Chaplin}, {Davies},
  {Garc{\'\i}a}, {Latham}, {Mathur}, {Mosser}, \& {Sharma}}]{Huber2017}
{Huber}, D., {Zinn}, J., {Bojsen-Hansen}, M., {et~al.} 2017, \apj, 844, 102,
  \dodoi{10.3847/1538-4357/aa75ca}

\bibitem[{{Hunter}(2007)}]{Hunter2007}
{Hunter}, J.~D. 2007, Computing in Science and Engineering, 9, 90,
  \dodoi{10.1109/MCSE.2007.55}

\bibitem[{{Hurt} \& {MacGregor}(2023)}]{Hurt2023}
{Hurt}, S.~A., \& {MacGregor}, M.~A. 2023, \apj, 954, 10,
  \dodoi{10.3847/1538-4357/accf9d}

\bibitem[{{Juri{\'c}} \& {Tremaine}(2008)}]{Juric2008}
{Juri{\'c}}, M., \& {Tremaine}, S. 2008, The Astrophysical Journal, 686, 603,
  \dodoi{10.1086/590047}

\bibitem[{{Kahraman Ali{\c{c}}avu{\c{s}}} {et~al.}(2016){Kahraman
  Ali{\c{c}}avu{\c{s}}}, {Niemczura}, {De Cat}, {Soydugan}, {Ko{\l}aczkowski},
  {Ostrowski}, {Telting}, {Uytterhoeven}, {Poretti}, {Rainer}, {Su{\'a}rez},
  {Mantegazza}, {Kilmartin}, \& {Pollard}}]{KahramanAlicavus2016}
{Kahraman Ali{\c{c}}avu{\c{s}}}, F., {Niemczura}, E., {De Cat}, P., {et~al.}
  2016, Monthly Notices of the Royal Astronomical Society, 458, 2307,
  \dodoi{10.1093/mnras/stw393}

\bibitem[{{Kaye} {et~al.}(1999){Kaye}, {Handler}, {Krisciunas}, {Poretti}, \&
  {Zerbi}}]{Kaye1999}
{Kaye}, A.~B., {Handler}, G., {Krisciunas}, K., {Poretti}, E., \& {Zerbi},
  F.~M. 1999, \pasp, 111, 840, \dodoi{10.1086/316399}

\bibitem[{{Kennedy} \& {Wyatt}(2010)}]{Kennedy2010}
{Kennedy}, G.~M., \& {Wyatt}, M.~C. 2010, Monthly Notices of the Royal
  Astronomical Society, 405, 1253, \dodoi{10.1111/j.1365-2966.2010.16528.x}

\bibitem[{{Kozai}(1962)}]{Kozai1962}
{Kozai}, Y. 1962, The Astronomical Journal, 67, 591, \dodoi{10.1086/108790}

\bibitem[{{Kraus} \& {Hillenbrand}(2009)}]{Kraus2009a}
{Kraus}, A.~L., \& {Hillenbrand}, L.~A. 2009, \apj, 704, 531,
  \dodoi{10.1088/0004-637X/704/1/531}

\bibitem[{{Krivov} \& {Booth}(2018)}]{Krivov2018}
{Krivov}, A.~V., \& {Booth}, M. 2018, Monthly Notices of the Royal Astronomical
  Society, 479, 3300, \dodoi{10.1093/mnras/sty1607}

\bibitem[{{Lai}(2012)}]{Lai2012}
{Lai}, D. 2012, Monthly Notices of the Royal Astronomical Society, 423, 486,
  \dodoi{10.1111/j.1365-2966.2012.20893.x}

\bibitem[{{Lai} {et~al.}(2011){Lai}, {Foucart}, \& {Lin}}]{Lai2011}
{Lai}, D., {Foucart}, F., \& {Lin}, D. N.~C. 2011, Monthly Notices of the Royal
  Astronomical Society, 412, 2790, \dodoi{10.1111/j.1365-2966.2010.18127.x}

\bibitem[{{Lammers} \& {Winn}(2026)}]{Lammers2026}
{Lammers}, C., \& {Winn}, J.~N. 2026, The Astronomical Journal, 171, 18,
  \dodoi{10.3847/1538-3881/ae21de}

\bibitem[{{Li} \& {Adams}(2015)}]{Li2015}
{Li}, G., \& {Adams}, F.~C. 2015, Monthly Notices of the Royal Astronomical
  Society, 448, 344, \dodoi{10.1093/mnras/stv012}

\bibitem[{{Lidov}(1962)}]{Lidov1962}
{Lidov}, M.~L. 1962, Planetary and Space Science, 9, 719,
  \dodoi{10.1016/0032-0633(62)90129-0}

\bibitem[{{Liebing} {et~al.}(2021){Liebing}, {Jeffers}, {Reiners}, \&
  {Zechmeister}}]{Liebing2021}
{Liebing}, F., {Jeffers}, S.~V., {Reiners}, A., \& {Zechmeister}, M. 2021,
  \aap, 654, A168, \dodoi{10.1051/0004-6361/202039607}

\bibitem[{{Lightkurve Collaboration} {et~al.}(2018){Lightkurve Collaboration},
  {Cardoso}, {Hedges}, {Gully-Santiago}, {Saunders}, {Cody}, {Barclay}, {Hall},
  {Sagear}, {Turtelboom}, {Zhang}, {Tzanidakis}, {Mighell}, {Coughlin}, {Bell},
  {Berta-Thompson}, {Williams}, {Dotson}, \&
  {Barentsen}}]{LightkurveCollaboration2018}
{Lightkurve Collaboration}, {Cardoso}, J. V. d.~M., {Hedges}, C., {et~al.}
  2018, {Lightkurve: Kepler and TESS time series analysis in Python},
  Astrophysics Source Code Library, record ascl:1812.013

\bibitem[{{Llorente de Andr{\'e}s} {et~al.}(2021){Llorente de Andr{\'e}s},
  {Chavero}, {de la Reza}, {Roca-F{\`a}brega}, \&
  {Cifuentes}}]{LlorentedeAndres2021}
{Llorente de Andr{\'e}s}, F., {Chavero}, C., {de la Reza}, R.,
  {Roca-F{\`a}brega}, S., \& {Cifuentes}, C. 2021, Astronomy and Astrophysics,
  654, A137, \dodoi{10.1051/0004-6361/202141339}

\bibitem[{{L{\'o}pez-Valdivia} {et~al.}(2019){L{\'o}pez-Valdivia}, {Mace},
  {Sokal}, {Hussaini}, {Kidder}, {Mann}, {Gosnell}, {Oh}, {Kesseli},
  {Muirhead}, {Johns-Krull}, \& {Jaffe}}]{Lopez-Valdivia2019}
{L{\'o}pez-Valdivia}, R., {Mace}, G.~N., {Sokal}, K.~R., {et~al.} 2019, \apj,
  879, 105, \dodoi{10.3847/1538-4357/ab2129}

\bibitem[{{L{\'o}pez-Valdivia} {et~al.}(2023){L{\'o}pez-Valdivia}, {Mace},
  {Han}, {Sawczynec}, {Hern{\'a}ndez}, {Prato}, {Johns-Krull}, {Oh}, {Lee},
  {Kraus}, {Llama}, \& {Jaffe}}]{Lopez-Valdivia2023}
{L{\'o}pez-Valdivia}, R., {Mace}, G.~N., {Han}, E., {et~al.} 2023, The
  Astrophysical Journal, 943, 49, \dodoi{10.3847/1538-4357/acab04}

\bibitem[{{Lovell} {et~al.}(2021){Lovell}, {Marino}, {Wyatt}, {Kennedy},
  {MacGregor}, {Stapelfeldt}, {Dent}, {Krist}, {Matr{\`a}}, {Kral},
  {Pani{\'c}}, {Pearce}, \& {Wilner}}]{Lovell2021}
{Lovell}, J.~B., {Marino}, S., {Wyatt}, M.~C., {et~al.} 2021, Monthly Notices
  of the Royal Astronomical Society, 506, 1978, \dodoi{10.1093/mnras/stab1678}

\bibitem[{{Loyd} \& {France}(2014)}]{Loyd2014}
{Loyd}, R.~O.~P., \& {France}, K. 2014, The Astrophysical Journal Supplement
  Series, 211, 9, \dodoi{10.1088/0067-0049/211/1/9}

\bibitem[{{Luck}(2017)}]{Luck2017}
{Luck}, R.~E. 2017, \aj, 153, 21, \dodoi{10.3847/1538-3881/153/1/21}

\bibitem[{{MacGregor} {et~al.}(2022){MacGregor}, {Hurt}, {Stark}, {Howard},
  {Weinberger}, {Ren}, {Schneider}, {Choquet}, \& {Mawet}}]{MacGregor2022}
{MacGregor}, M.~A., {Hurt}, S.~A., {Stark}, C.~C., {et~al.} 2022, The
  Astrophysical Journal, 933, L1, \dodoi{10.3847/2041-8213/ac7729}

\bibitem[{{Macintosh} {et~al.}(2014){Macintosh}, {Graham}, {Ingraham},
  {Konopacky}, {Marois}, {Perrin}, {Poyneer}, {Bauman}, {Barman}, {Burrows},
  {Cardwell}, {Chilcote}, {De Rosa}, {Dillon}, {Doyon}, {Dunn}, {Erikson},
  {Fitzgerald}, {Gavel}, {Goodsell}, {Hartung}, {Hibon}, {Kalas}, {Larkin},
  {Maire}, {Marchis}, {Marley}, {McBride}, {Millar-Blanchaer}, {Morzinski},
  {Norton}, {Oppenheimer}, {Palmer}, {Patience}, {Pueyo}, {Rantakyro},
  {Sadakuni}, {Saddlemyer}, {Savransky}, {Serio}, {Soummer},
  {Sivaramakrishnan}, {Song}, {Thomas}, {Wallace}, {Wiktorowicz}, \&
  {Wolff}}]{Macintosh2014}
{Macintosh}, B., {Graham}, J.~R., {Ingraham}, P., {et~al.} 2014, Proceedings of
  the National Academy of Science, 111, 12661, \dodoi{10.1073/pnas.1304215111}

\bibitem[{{Maldonado} {et~al.}(2015){Maldonado}, {Eiroa}, {Villaver},
  {Montesinos}, \& {Mora}}]{Maldonado2015}
{Maldonado}, J., {Eiroa}, C., {Villaver}, E., {Montesinos}, B., \& {Mora}, A.
  2015, Astronomy and Astrophysics, 579, A20,
  \dodoi{10.1051/0004-6361/201525764}

\bibitem[{{Maldonado} {et~al.}(2010){Maldonado}, {Mart{\'\i}nez-Arn{\'a}iz},
  {Eiroa}, {Montes}, \& {Montesinos}}]{Maldonado2010}
{Maldonado}, J., {Mart{\'\i}nez-Arn{\'a}iz}, R.~M., {Eiroa}, C., {Montes}, D.,
  \& {Montesinos}, B. 2010, Astronomy and Astrophysics, 521, A12,
  \dodoi{10.1051/0004-6361/201014948}

\bibitem[{{Malmberg} {et~al.}(2011){Malmberg}, {Davies}, \&
  {Heggie}}]{Malmberg2011}
{Malmberg}, D., {Davies}, M.~B., \& {Heggie}, D.~C. 2011, Monthly Notices of
  the Royal Astronomical Society, 411, 859,
  \dodoi{10.1111/j.1365-2966.2010.17730.x}

\bibitem[{{Malo} {et~al.}(2014){Malo}, {Artigau}, {Doyon}, {Lafreni{\`e}re},
  {Albert}, \& {Gagn{\'e}}}]{Malo2014}
{Malo}, L., {Artigau}, {\'E}., {Doyon}, R., {et~al.} 2014, \apj, 788, 81,
  \dodoi{10.1088/0004-637X/788/1/81}

\bibitem[{{Mamajek} {et~al.}(2013){Mamajek}, {Bartlett}, {Seifahrt}, {Henry},
  {Dieterich}, {Lurie}, {Kenworthy}, {Jao}, {Riedel}, {Subasavage}, {Winters},
  {Finch}, {Ianna}, \& {Bean}}]{Mamajek2013}
{Mamajek}, E.~E., {Bartlett}, J.~L., {Seifahrt}, A., {et~al.} 2013, The
  Astronomical Journal, 146, 154, \dodoi{10.1088/0004-6256/146/6/154}

\bibitem[{{Marfil} {et~al.}(2021){Marfil}, {Tabernero}, {Montes}, {Caballero},
  {L{\'a}zaro}, {Gonz{\'a}lez Hern{\'a}ndez}, {Nagel}, {Passegger},
  {Schweitzer}, {Ribas}, {Reiners}, {Quirrenbach}, {Amado}, {Cifuentes},
  {Cort{\'e}s-Contreras}, {Dreizler}, {Duque-Arribas},
  {Galad{\'\i}-Enr{\'\i}quez}, {Henning}, {Jeffers}, {Kaminski}, {K{\"u}rster},
  {Lafarga}, {L{\'o}pez-Gallifa}, {Morales}, {Shan}, \&
  {Zechmeister}}]{Marfil2021}
{Marfil}, E., {Tabernero}, H.~M., {Montes}, D., {et~al.} 2021, Astronomy and
  Astrophysics, 656, A162, \dodoi{10.1051/0004-6361/202141980}

\bibitem[{{Marino} {et~al.}(2019){Marino}, {Yelverton}, {Booth}, {Faramaz},
  {Kennedy}, {Matr{\`a}}, \& {Wyatt}}]{Marino2019}
{Marino}, S., {Yelverton}, B., {Booth}, M., {et~al.} 2019, Monthly Notices of
  the Royal Astronomical Society, 484, 1257, \dodoi{10.1093/mnras/stz049}

\bibitem[{{Marsden} {et~al.}(2014){Marsden}, {Petit}, {Jeffers}, {Morin},
  {Fares}, {Reiners}, {do Nascimento}, {Auri{\`e}re}, {Bouvier}, {Carter},
  {Catala}, {Dintrans}, {Donati}, {Gastine}, {Jardine}, {Konstantinova-Antova},
  {Lanoux}, {Ligni{\`e}res}, {Morgenthaler}, {Ram{\`\i}rez-V{\`e}lez},
  {Th{\'e}ado}, {Van Grootel}, \& {BCool Collaboration}}]{Marsden2014}
{Marsden}, S.~C., {Petit}, P., {Jeffers}, S.~V., {et~al.} 2014, Monthly Notices
  of the Royal Astronomical Society, 444, 3517, \dodoi{10.1093/mnras/stu1663}

\bibitem[{{Marshall} {et~al.}(2026){Marshall}, {Mu{\~n}oz-Guti{\'e}rrez},
  {Sefilian}, \& {Peimbert}}]{Marshall2026}
{Marshall}, J.~P., {Mu{\~n}oz-Guti{\'e}rrez}, M.~A., {Sefilian}, A.~A., \&
  {Peimbert}, A. 2026, Monthly Notices of the Royal Astronomical Society, 545,
  staf1990, \dodoi{10.1093/mnras/staf1990}

\bibitem[{{Marshall} {et~al.}(2021){Marshall}, {Wang}, {Kennedy}, {Zeegers}, \&
  {Scicluna}}]{Marshall2021}
{Marshall}, J.~P., {Wang}, L., {Kennedy}, G.~M., {Zeegers}, S.~T., \&
  {Scicluna}, P. 2021, Monthly Notices of the Royal Astronomical Society, 501,
  6168, \dodoi{10.1093/mnras/staa3917}

\bibitem[{{Mart{\'\i}nez-Arn{\'a}iz} {et~al.}(2010){Mart{\'\i}nez-Arn{\'a}iz},
  {Maldonado}, {Montes}, {Eiroa}, \& {Montesinos}}]{Martinez-Arnaiz2010}
{Mart{\'\i}nez-Arn{\'a}iz}, R., {Maldonado}, J., {Montes}, D., {Eiroa}, C., \&
  {Montesinos}, B. 2010, Astronomy and Astrophysics, 520, A79,
  \dodoi{10.1051/0004-6361/200913725}

\bibitem[{{Masuda} \& {Winn}(2020)}]{Masuda2020}
{Masuda}, K., \& {Winn}, J.~N. 2020, \aj, 159, 81,
  \dodoi{10.3847/1538-3881/ab65be}

\bibitem[{{Matr{\`a}} {et~al.}(2025){Matr{\`a}}, {Marino}, {Wilner}, {Kennedy},
  {Booth}, {Krivov}, {Williams}, {Hughes}, {del Burgo}, {Carpenter}, {Davies},
  {Ertel}, {Kral}, {Lestrade}, {Marshall}, {Milli}, {{\"O}berg}, {Pawellek},
  {Sepulveda}, {Wyatt}, {Matthews}, \& {MacGregor}}]{Matra2025}
{Matr{\`a}}, L., {Marino}, S., {Wilner}, D.~J., {et~al.} 2025, Astronomy and
  Astrophysics, 693, A151, \dodoi{10.1051/0004-6361/202451397}

\bibitem[{{McArthur} {et~al.}(2010){McArthur}, {Benedict}, {Barnes},
  {Martioli}, {Korzennik}, {Nelan}, \& {Butler}}]{McArthur2010}
{McArthur}, B.~E., {Benedict}, G.~F., {Barnes}, R., {et~al.} 2010, The
  Astrophysical Journal, 715, 1203, \dodoi{10.1088/0004-637X/715/2/1203}

\bibitem[{{McLaughlin}(1924)}]{McLaughlin1924}
{McLaughlin}, D.~B. 1924, \apj, 60, 22, \dodoi{10.1086/142826}

\bibitem[{{Mellon} {et~al.}(2017){Mellon}, {Mamajek}, {Oberst}, \&
  {Pecaut}}]{Mellon2017}
{Mellon}, S.~N., {Mamajek}, E.~E., {Oberst}, T.~E., \& {Pecaut}, M.~J. 2017,
  \apj, 844, 66, \dodoi{10.3847/1538-4357/aa77fb}

\bibitem[{{Mentuch} {et~al.}(2008){Mentuch}, {Brandeker}, {van Kerkwijk},
  {Jayawardhana}, \& {Hauschildt}}]{Mentuch2008}
{Mentuch}, E., {Brandeker}, A., {van Kerkwijk}, M.~H., {Jayawardhana}, R., \&
  {Hauschildt}, P.~H. 2008, The Astrophysical Journal, 689, 1127,
  \dodoi{10.1086/592764}

\bibitem[{{Messina} {et~al.}(2010){Messina}, {Desidera}, {Turatto},
  {Lanzafame}, \& {Guinan}}]{Messina2010}
{Messina}, S., {Desidera}, S., {Turatto}, M., {Lanzafame}, A.~C., \& {Guinan},
  E.~F. 2010, \aap, 520, A15, \dodoi{10.1051/0004-6361/200913644}

\bibitem[{{Mills} \& {Fabrycky}(2017)}]{Mills2017}
{Mills}, S.~M., \& {Fabrycky}, D.~C. 2017, The Astrophysical Journal, 838, L11,
  \dodoi{10.3847/2041-8213/aa6543}

\bibitem[{{Moe} \& {Kratter}(2021)}]{Moe2021}
{Moe}, M., \& {Kratter}, K.~M. 2021, Monthly Notices of the Royal Astronomical
  Society, 507, 3593, \dodoi{10.1093/mnras/stab2328}

\bibitem[{{Montalto} {et~al.}(2021){Montalto}, {Piotto}, {Marrese},
  {Nascimbeni}, {Prisinzano}, {Granata}, {Marinoni}, {Desidera}, {Ortolani},
  {Aerts}, {Alei}, {Altavilla}, {Benatti}, {B{\"o}rner}, {Cabrera}, {Claudi},
  {Deleuil}, {Fabrizio}, {Gizon}, {Goupil}, {Heras}, {Magrin}, {Malavolta},
  {Mas-Hesse}, {Pagano}, {Paproth}, {Pertenais}, {Pollacco}, {Ragazzoni},
  {Ramsay}, {Rauer}, \& {Udry}}]{Montalto2021}
{Montalto}, M., {Piotto}, G., {Marrese}, P.~M., {et~al.} 2021, \aap, 653, A98,
  \dodoi{10.1051/0004-6361/202140717}

\bibitem[{{Morgan} {et~al.}(2026){Morgan}, {Bowler}, \& {Tran}}]{Morgan2026}
{Morgan}, M., {Bowler}, B.~P., \& {Tran}, Q.~H. 2026, The Astronomical Journal,
  171, 19, \dodoi{10.3847/1538-3881/ae0e16}

\bibitem[{{Morgan} {et~al.}(2024){Morgan}, {Bowler}, {Tran}, {Petigura},
  {Nagpal}, \& {Blunt}}]{Morgan2024}
{Morgan}, M., {Bowler}, B.~P., {Tran}, Q.~H., {et~al.} 2024, \aj, 167, 48,
  \dodoi{10.3847/1538-3881/ad0728}

\bibitem[{{Morgan} {et~al.}(2025){Morgan}, {Bowler}, {Tran}, {Wittenmyer},
  {Wright}, {Zhou}, \& {Fairnington}}]{Morgan2025}
---. 2025, The Astrophysical Journal Supplement Series, 280, 76,
  \dodoi{10.3847/1538-4365/adf55d}

\bibitem[{{Mortier} {et~al.}(2013){Mortier}, {Santos}, {Sousa}, {Israelian},
  {Mayor}, \& {Udry}}]{Mortier2013}
{Mortier}, A., {Santos}, N.~C., {Sousa}, S., {et~al.} 2013, Astronomy and
  Astrophysics, 551, A112, \dodoi{10.1051/0004-6361/201220707}

\bibitem[{{Mustill} \& {Wyatt}(2009)}]{Mustill2009}
{Mustill}, A.~J., \& {Wyatt}, M.~C. 2009, Monthly Notices of the Royal
  Astronomical Society, 399, 1403, \dodoi{10.1111/j.1365-2966.2009.15360.x}

\bibitem[{{Neveu-VanMalle} {et~al.}(2016){Neveu-VanMalle}, {Queloz},
  {Anderson}, {Brown}, {Collier Cameron}, {Delrez}, {D{\'\i}az}, {Gillon},
  {Hellier}, {Jehin}, {Lister}, {Pepe}, {Rojo}, {S{\'e}gransan}, {Triaud},
  {Turner}, \& {Udry}}]{Neveu-VanMalle2016}
{Neveu-VanMalle}, M., {Queloz}, D., {Anderson}, D.~R., {et~al.} 2016, Astronomy
  and Astrophysics, 586, A93, \dodoi{10.1051/0004-6361/201526965}

\bibitem[{{Nicholson} {et~al.}(2021){Nicholson}, {Hussain}, {Donati}, {Wright},
  {Folsom}, {Wittenmyer}, {Okumura}, \& {Carter}}]{Nicholson2021}
{Nicholson}, B.~A., {Hussain}, G., {Donati}, J.~F., {et~al.} 2021, Monthly
  Notices of the Royal Astronomical Society, 504, 2461,
  \dodoi{10.1093/mnras/stab879}

\bibitem[{{Nordstr{\"o}m} {et~al.}(2004){Nordstr{\"o}m}, {Mayor}, {Andersen},
  {Holmberg}, {Pont}, {J{\o}rgensen}, {Olsen}, {Udry}, \&
  {Mowlavi}}]{Nordstrom2004}
{Nordstr{\"o}m}, B., {Mayor}, M., {Andersen}, J., {et~al.} 2004, Astronomy and
  Astrophysics, 418, 989, \dodoi{10.1051/0004-6361:20035959}

\bibitem[{{Noyes} {et~al.}(1984){Noyes}, {Hartmann}, {Baliunas}, {Duncan}, \&
  {Vaughan}}]{Noyes1984}
{Noyes}, R.~W., {Hartmann}, L.~W., {Baliunas}, S.~L., {Duncan}, D.~K., \&
  {Vaughan}, A.~H. 1984, The Astrophysical Journal, 279, 763,
  \dodoi{10.1086/161945}

\bibitem[{Ochsenbein(1996)}]{10.26093/cds/vizier}
Ochsenbein, F. 1996, The VizieR database of astronomical catalogues,  CDS,
  Centre de DonnÃ©es astronomiques de Strasbourg, \dodoi{10.26093/CDS/VIZIER}

\bibitem[{{Ochsenbein} {et~al.}(2000){Ochsenbein}, {Bauer}, \&
  {Marcout}}]{vizier2000}
{Ochsenbein}, F., {Bauer}, P., \& {Marcout}, J. 2000, \aaps, 143, 23,
  \dodoi{10.1051/aas:2000169}

\bibitem[{{Palatnick} {et~al.}(2025){Palatnick}, {Millar-Blanchaer}, {Zhang},
  {Lawson}, {Lewis}, {Crotts}, {Carter}, {Biller}, {Girard}, {Marino},
  {Bendahan-West}, {Strampelli}, {James}, {Stephenson}, {Ferrer-Chavez},
  {Booth}, {Sutlieff}, {Sanghi}, {Fontanive}, {Rickman}, {Rebollido}, {Hoch},
  \& {Balmer}}]{Palatnick2025}
{Palatnick}, S., {Millar-Blanchaer}, M.~A., {Zhang}, J., {et~al.} 2025, The
  Astrophysical Journal, 995, 149, \dodoi{10.3847/1538-4357/ae1963}

\bibitem[{{Pass} {et~al.}(2023){Pass}, {Winters}, {Charbonneau}, {Irwin}, \&
  {Medina}}]{Pass2023}
{Pass}, E.~K., {Winters}, J.~G., {Charbonneau}, D., {Irwin}, J.~M., \&
  {Medina}, A.~A. 2023, The Astronomical Journal, 166, 16,
  \dodoi{10.3847/1538-3881/acd6a2}

\bibitem[{{Pearce} {et~al.}(2022){Pearce}, {Launhardt}, {Ostermann}, {Kennedy},
  {Gennaro}, {Booth}, {Krivov}, {Cugno}, {Henning}, {Quirrenbach}, {Barcucci},
  {Matthews}, {Ruh}, \& {Stone}}]{Pearce2022}
{Pearce}, T.~D., {Launhardt}, R., {Ostermann}, R., {et~al.} 2022, Astronomy and
  Astrophysics, 659, A135, \dodoi{10.1051/0004-6361/202142720}

\bibitem[{{Perdelwitz} {et~al.}(2024){Perdelwitz}, {Trifonov}, {Teklu},
  {Sreenivas}, \& {Tal-Or}}]{Perdelwitz2024}
{Perdelwitz}, V., {Trifonov}, T., {Teklu}, J.~T., {Sreenivas}, K.~R., \&
  {Tal-Or}, L. 2024, Astronomy and Astrophysics, 683, A125,
  \dodoi{10.1051/0004-6361/202348263}

\bibitem[{{Perryman} {et~al.}(2014){Perryman}, {Hartman}, {Bakos}, \&
  {Lindegren}}]{Perryman2014}
{Perryman}, M., {Hartman}, J., {Bakos}, G.~{\'A}., \& {Lindegren}, L. 2014, The
  Astrophysical Journal, 797, 14, \dodoi{10.1088/0004-637X/797/1/14}

\bibitem[{{Pilbratt} {et~al.}(2010){Pilbratt}, {Riedinger}, {Passvogel},
  {Crone}, {Doyle}, {Gageur}, {Heras}, {Jewell}, {Metcalfe}, {Ott}, \&
  {Schmidt}}]{Pilbratt2010}
{Pilbratt}, G.~L., {Riedinger}, J.~R., {Passvogel}, T., {et~al.} 2010,
  Astronomy and Astrophysics, 518, L1, \dodoi{10.1051/0004-6361/201014759}

\bibitem[{{Pinsonneault} {et~al.}(2001){Pinsonneault}, {DePoy}, \&
  {Coffee}}]{Pinsonneault2001}
{Pinsonneault}, M.~H., {DePoy}, D.~L., \& {Coffee}, M. 2001, The Astrophysical
  Journal, 556, L59, \dodoi{10.1086/323531}

\bibitem[{{Pizzolato} {et~al.}(2003){Pizzolato}, {Maggio}, {Micela},
  {Sciortino}, \& {Ventura}}]{Pizzolato2003}
{Pizzolato}, N., {Maggio}, A., {Micela}, G., {Sciortino}, S., \& {Ventura}, P.
  2003, Astronomy and Astrophysics, 397, 147,
  \dodoi{10.1051/0004-6361:20021560}

\bibitem[{{Plavchan} {et~al.}(2009){Plavchan}, {Werner}, {Chen}, {Stapelfeldt},
  {Su}, {Stauffer}, \& {Song}}]{Plavchan2009}
{Plavchan}, P., {Werner}, M.~W., {Chen}, C.~H., {et~al.} 2009, The
  Astrophysical Journal, 698, 1068, \dodoi{10.1088/0004-637X/698/2/1068}

\bibitem[{{Raymond} {et~al.}(2011){Raymond}, {Armitage}, {Moro-Mart{\'\i}n},
  {Booth}, {Wyatt}, {Armstrong}, {Mandell}, {Selsis}, \& {West}}]{Raymond2011}
{Raymond}, S.~N., {Armitage}, P.~J., {Moro-Mart{\'\i}n}, A., {et~al.} 2011,
  Astronomy and Astrophysics, 530, A62, \dodoi{10.1051/0004-6361/201116456}

\bibitem[{{Reid}(2003)}]{Reid2003}
{Reid}, N. 2003, Monthly Notices of the Royal Astronomical Society, 342, 837,
  \dodoi{10.1046/j.1365-8711.2003.06581.x}

\bibitem[{{Reiners} {et~al.}(2022){Reiners}, {Shulyak}, {K{\"a}pyl{\"a}},
  {Ribas}, {Nagel}, {Zechmeister}, {Caballero}, {Shan}, {Fuhrmeister},
  {Quirrenbach}, {Amado}, {Montes}, {Jeffers}, {Azzaro}, {B{\'e}jar},
  {Chaturvedi}, {Henning}, {K{\"u}rster}, \& {Pall{\'e}}}]{Reiners2022}
{Reiners}, A., {Shulyak}, D., {K{\"a}pyl{\"a}}, P.~J., {et~al.} 2022, Astronomy
  and Astrophysics, 662, A41, \dodoi{10.1051/0004-6361/202243251}

\bibitem[{{Ren} {et~al.}(2021){Ren}, {Choquet}, {Perrin}, {Mawet}, {Chen},
  {Milli}, {Debes}, {Rebollido}, {Stark}, {Hagan}, {Hines}, {Millar-Blanchaer},
  {Pueyo}, {Roberge}, {Schneider}, {Serabyn}, {Soummer}, \& {Wolff}}]{Ren2021}
{Ren}, B., {Choquet}, {\'E}., {Perrin}, M.~D., {et~al.} 2021, The Astrophysical
  Journal, 914, 95, \dodoi{10.3847/1538-4357/ac03b9}

\bibitem[{{Rice} {et~al.}(2022){Rice}, {Wang}, {Wang}, {Stef{\'a}nsson},
  {Isaacson}, {Howard}, {Logsdon}, {Schweiker}, {Dai}, {Brinkman}, {Giacalone},
  \& {Holcomb}}]{Rice2022a}
{Rice}, M., {Wang}, S., {Wang}, X.-Y., {et~al.} 2022, \aj, 164, 104,
  \dodoi{10.3847/1538-3881/ac8153}

\bibitem[{{Robitaille} {et~al.}(2021){Robitaille}, {Tollerud}, {Aldcroft},
  {Bray}, {Van Kerkwijk}, {Droettboom}, {Sip{\H{o}}cz}, {Lim}, {Price-Whelan},
  {Conseil}, {Dencheva}, {Bradley}, {Ginsburg}, {Mumford}, {Seifert}, {Craig},
  {Mdmueller}, {StuartLittlefair}, {D'Avella}, {Lglattly}, {G{\"u}nther},
  {Homeier}, {Donath}, {Perrygreenfield}, {Deil}, {Vin{\'\i}cius}, {Woillez},
  {Vanderplas}, {Patil}, \& {N{\"o}the}}]{Robitaille2021}
{Robitaille}, T., {Tollerud}, E., {Aldcroft}, T., {et~al.} 2021,
  {astropy/astropy: v4.2.1}, v4.2.1,  Zenodo, \dodoi{10.5281/zenodo.4670729}

\bibitem[{{Rogers} \& {Lin}(2013)}]{Rogers2013}
{Rogers}, T.~M., \& {Lin}, D.~N.~C. 2013, The Astrophysical Journal, 769, L10,
  \dodoi{10.1088/2041-8205/769/1/L10}

\bibitem[{{Rossiter}(1924)}]{Rossiter1924}
{Rossiter}, R.~A. 1924, \apj, 60, 15, \dodoi{10.1086/142825}

\bibitem[{{Rusznak} {et~al.}(2025){Rusznak}, {Wang}, {Rice}, \&
  {Wang}}]{Rusznak2025}
{Rusznak}, J., {Wang}, X.-Y., {Rice}, M., \& {Wang}, S. 2025, The Astrophysical
  Journal, 983, L42, \dodoi{10.3847/2041-8213/adc129}

\bibitem[{{Saar} \& {Osten}(1997)}]{Saar1997}
{Saar}, S.~H., \& {Osten}, R.~A. 1997, Monthly Notices of the Royal
  Astronomical Society, 284, 803, \dodoi{10.1093/mnras/284.4.803}

\bibitem[{{Sarmento} {et~al.}(2020){Sarmento}, {Delgado Mena}, {Rojas-Ayala},
  \& {Blanco-Cuaresma}}]{Sarmento2020}
{Sarmento}, P., {Delgado Mena}, E., {Rojas-Ayala}, B., \& {Blanco-Cuaresma}, S.
  2020, Astronomy and Astrophysics, 636, A85,
  \dodoi{10.1051/0004-6361/201936296}

\bibitem[{{Schlaufman}(2010)}]{Schlaufman2010}
{Schlaufman}, K.~C. 2010, \apj, 719, 602, \dodoi{10.1088/0004-637X/719/1/602}

\bibitem[{{Schmidt} \& {Schlaufman}(2026)}]{Schmidt2026}
{Schmidt}, S.~P., \& {Schlaufman}, K.~C. 2026, The Astronomical Journal, 171,
  157, \dodoi{10.3847/1538-3881/ae3c11}

\bibitem[{{Schneider} {et~al.}(2012){Schneider}, {Melis}, \&
  {Song}}]{Schneider2012}
{Schneider}, A., {Melis}, C., \& {Song}, I. 2012, The Astrophysical Journal,
  754, 39, \dodoi{10.1088/0004-637X/754/1/39}

\bibitem[{{Schneider} {et~al.}(2016){Schneider}, {Grady}, {Stark}, {Gaspar},
  {Carson}, {Debes}, {Henning}, {Hines}, {Jang-Condell}, {Kuchner}, {Perrin},
  {Rodigas}, {Tamura}, \& {Wisniewski}}]{Schneider2016}
{Schneider}, G., {Grady}, C.~A., {Stark}, C.~C., {et~al.} 2016, The
  Astronomical Journal, 152, 64, \dodoi{10.3847/0004-6256/152/3/64}

\bibitem[{{Schr{\"o}der} {et~al.}(2009){Schr{\"o}der}, {Reiners}, \&
  {Schmitt}}]{Schroder2009}
{Schr{\"o}der}, C., {Reiners}, A., \& {Schmitt}, J.~H.~M.~M. 2009, Astronomy
  and Astrophysics, 493, 1099, \dodoi{10.1051/0004-6361:200810377}

\bibitem[{{Sepulveda} {et~al.}(2024){Sepulveda}, {Huber}, {Bedding}, {Hey},
  {Murphy}, {Zhang}, \& {Liu}}]{Sepulveda2024a}
{Sepulveda}, A.~G., {Huber}, D., {Bedding}, T.~R., {et~al.} 2024, \aj, 168, 13,
  \dodoi{10.3847/1538-3881/ad4964}

\bibitem[{{Sepulveda} {et~al.}(2022){Sepulveda}, {Huber}, {Zhang}, {Li}, {Liu},
  \& {Bedding}}]{Sepulveda2022}
{Sepulveda}, A.~G., {Huber}, D., {Zhang}, Z., {et~al.} 2022, \apj, 938, 49,
  \dodoi{10.3847/1538-4357/ac9229}

\bibitem[{{Song} {et~al.}(2004){Song}, {Zuckerman}, \& {Bessell}}]{Song2004}
{Song}, I., {Zuckerman}, B., \& {Bessell}, M.~S. 2004, The Astrophysical
  Journal, 614, L125, \dodoi{10.1086/425683}

\bibitem[{{Soto} \& {Jenkins}(2018)}]{Soto2018}
{Soto}, M.~G., \& {Jenkins}, J.~S. 2018, Astronomy and Astrophysics, 615, A76,
  \dodoi{10.1051/0004-6361/201731533}

\bibitem[{{Soubiran} {et~al.}(2022){Soubiran}, {Brouillet}, \&
  {Casamiquela}}]{Soubiran2022}
{Soubiran}, C., {Brouillet}, N., \& {Casamiquela}, L. 2022, Astronomy and
  Astrophysics, 663, A4, \dodoi{10.1051/0004-6361/202142409}

\bibitem[{{Southworth}(2026)}]{Southworth2026}
{Southworth}, J. 2026, Universe, 12, 62, \dodoi{10.3390/universe12030062}

\bibitem[{{Spalding} \& {Batygin}(2014)}]{Spalding2014}
{Spalding}, C., \& {Batygin}, K. 2014, The Astrophysical Journal, 790, 42,
  \dodoi{10.1088/0004-637X/790/1/42}

\bibitem[{{Spangler} {et~al.}(2001){Spangler}, {Sargent}, {Silverstone},
  {Becklin}, \& {Zuckerman}}]{Spangler2001}
{Spangler}, C., {Sargent}, A.~I., {Silverstone}, M.~D., {Becklin}, E.~E., \&
  {Zuckerman}, B. 2001, \apj, 555, 932, \dodoi{10.1086/321490}

\bibitem[{{Stassun} {et~al.}(2019){Stassun}, {Oelkers}, {Paegert}, {Torres},
  {Pepper}, {De Lee}, {Collins}, {Latham}, {Muirhead}, {Chittidi},
  {Rojas-Ayala}, {Fleming}, {Rose}, {Tenenbaum}, {Ting}, {Kane}, {Barclay},
  {Bean}, {Brassuer}, {Charbonneau}, {Ge}, {Lissauer}, {Mann}, {McLean},
  {Mullally}, {Narita}, {Plavchan}, {Ricker}, {Sasselov}, {Seager}, {Sharma},
  {Shiao}, {Sozzetti}, {Stello}, {Vanderspek}, {Wallace}, \&
  {Winn}}]{Stassun2019}
{Stassun}, K.~G., {Oelkers}, R.~J., {Paegert}, M., {et~al.} 2019, \aj, 158,
  138, \dodoi{10.3847/1538-3881/ab3467}

\bibitem[{{Stevens} {et~al.}(2017){Stevens}, {Stassun}, \&
  {Gaudi}}]{Stevens2017}
{Stevens}, D.~J., {Stassun}, K.~G., \& {Gaudi}, B.~S. 2017, The Astronomical
  Journal, 154, 259, \dodoi{10.3847/1538-3881/aa957b}

\bibitem[{{Storch} {et~al.}(2014){Storch}, {Anderson}, \& {Lai}}]{Storch2014}
{Storch}, N.~I., {Anderson}, K.~R., \& {Lai}, D. 2014, Science, 345, 1317,
  \dodoi{10.1126/science.1254358}

\bibitem[{{Tabernero} {et~al.}(2024){Tabernero}, {Shan}, {Caballero},
  {Duque-Arribas}, {Montes}, {Gonz{\'a}lez Hern{\'a}ndez}, {Zapatero Osorio},
  {Schweitzer}, {Henning}, {Cort{\'e}s-Contreras}, {Quirrenbach}, {Amado},
  {Reiners}, {Ribas}, {Bergond}, \& {Morales}}]{Tabernero2024}
{Tabernero}, H.~M., {Shan}, Y., {Caballero}, J.~A., {et~al.} 2024, Astronomy
  and Astrophysics, 689, A223, \dodoi{10.1051/0004-6361/202450054}

\bibitem[{{Takeda} {et~al.}(2010){Takeda}, {Honda}, {Kawanomoto}, {Ando}, \&
  {Sakurai}}]{Takeda2010}
{Takeda}, Y., {Honda}, S., {Kawanomoto}, S., {Ando}, H., \& {Sakurai}, T. 2010,
  Astronomy and Astrophysics, 515, A93, \dodoi{10.1051/0004-6361/200913897}

\bibitem[{{Takeda} {et~al.}(2005){Takeda}, {Sato}, {Kambe}, {Masuda},
  {Izumiura}, {Watanabe}, {Ohkubo}, {Yanagisawa}, {Yasaka}, {Honda},
  {Kawanomoto}, {Aoki}, {Ando}, {Sadakane}, {Takada-Hidai}, \&
  {Miyama}}]{Takeda2005}
{Takeda}, Y., {Sato}, B., {Kambe}, E., {et~al.} 2005, Publications of the
  Astronomical Society of Japan, 57, 13, \dodoi{10.1093/pasj/57.1.13}

\bibitem[{{Torres} {et~al.}(2006){Torres}, {Quast}, {da Silva}, {de La Reza},
  {Melo}, \& {Sterzik}}]{Torres2006}
{Torres}, C.~A.~O., {Quast}, G.~R., {da Silva}, L., {et~al.} 2006, \aap, 460,
  695, \dodoi{10.1051/0004-6361:20065602}

\bibitem[{{Trifonov} {et~al.}(2020){Trifonov}, {Tal-Or}, {Zechmeister},
  {Kaminski}, {Zucker}, \& {Mazeh}}]{Trifonov2020}
{Trifonov}, T., {Tal-Or}, L., {Zechmeister}, M., {et~al.} 2020, \aap, 636, A74,
  \dodoi{10.1051/0004-6361/201936686}

\bibitem[{{Trilling} {et~al.}(2007){Trilling}, {Stansberry}, {Stapelfeldt},
  {Rieke}, {Su}, {Gray}, {Corbally}, {Bryden}, {Chen}, {Boden}, \&
  {Beichman}}]{Trilling2007}
{Trilling}, D.~E., {Stansberry}, J.~A., {Stapelfeldt}, K.~R., {et~al.} 2007,
  The Astrophysical Journal, 658, 1289, \dodoi{10.1086/511668}

\bibitem[{{Valenti} \& {Fischer}(2005)}]{Valenti2005}
{Valenti}, J.~A., \& {Fischer}, D.~A. 2005, \apjs, 159, 141,
  \dodoi{10.1086/430500}

\bibitem[{{van der Walt} {et~al.}(2011){van der Walt}, {Colbert}, \&
  {Varoquaux}}]{vanderWalt2011}
{van der Walt}, S., {Colbert}, S.~C., \& {Varoquaux}, G. 2011, Computing in
  Science and Engineering, 13, 22, \dodoi{10.1109/MCSE.2011.37}

\bibitem[{{VanderPlas}(2018)}]{VanderPlas2018}
{VanderPlas}, J.~T. 2018, \apjs, 236, 16, \dodoi{10.3847/1538-4365/aab766}

\bibitem[{{Vaughan} {et~al.}(1981){Vaughan}, {Baliunas}, {Middelkoop},
  {Hartmann}, {Mihalas}, {Noyes}, \& {Preston}}]{Vaughan1981}
{Vaughan}, A.~H., {Baliunas}, S.~L., {Middelkoop}, F., {et~al.} 1981, The
  Astrophysical Journal, 250, 276, \dodoi{10.1086/159372}

\bibitem[{{Vincent} {et~al.}(2024){Vincent}, {Lawson}, {Currie}, {Williams},
  {Guyon}, {Lozi}, {Deo}, \& {Vievard}}]{Vincent2024}
{Vincent}, M., {Lawson}, K., {Currie}, T., {et~al.} 2024, The Astronomical
  Journal, 168, 6, \dodoi{10.3847/1538-3881/ad4864}

\bibitem[{{Virtanen} {et~al.}(2021){Virtanen}, {Gommers}, {Burovski},
  {Oliphant}, {Weckesser}, {Cournapeau}, {Alexbrc}, {Reddy}, {Peterson},
  {Haberland}, {Wilson}, {Nelson}, {Endolith}, {Mayorov}, {Van Der Walt},
  {Polat}, {Laxalde}, {Brett}, {Larson}, {Millman}, {Lars}, {Van Mulbregt},
  {Eric-Jones}, {Carey}, {Peterbell10}, {Moore}, {Kern}, {Leslie}, {Perktold},
  \& {Striega}}]{Virtanen2021}
{Virtanen}, P., {Gommers}, R., {Burovski}, E., {et~al.} 2021, {scipy/scipy:
  SciPy 1.6.3}, v1.6.3,  Zenodo, \dodoi{10.5281/zenodo.4718897}

\bibitem[{{von Zeipel}(1910)}]{vonZeipel1910}
{von Zeipel}, H. 1910, Astronomische Nachrichten, 183, 345,
  \dodoi{10.1002/asna.19091832202}

\bibitem[{{Wang} {et~al.}(2024){Wang}, {Rice}, {Wang}, {Kanodia}, {Dai},
  {Logsdon}, {Schweiker}, {Teske}, {Butler}, {Crane}, {Shectman}, {Quinn},
  {Kostov}, {Osborn}, {Goeke}, {Eastman}, {Shporer}, {Rapetti}, {Collins},
  {Watkins}, {Relles}, {Ricker}, {Seager}, {Winn}, \& {Jenkins}}]{Wang2024}
{Wang}, X.-Y., {Rice}, M., {Wang}, S., {et~al.} 2024, The Astrophysical
  Journal, 973, L21, \dodoi{10.3847/2041-8213/ad7469}

\bibitem[{{Weise} {et~al.}(2010){Weise}, {Launhardt}, {Setiawan}, \&
  {Henning}}]{Weise2010}
{Weise}, P., {Launhardt}, R., {Setiawan}, J., \& {Henning}, T. 2010, \aap, 517,
  A88, \dodoi{10.1051/0004-6361/201014453}

\bibitem[{{White} {et~al.}(2007){White}, {Greene}, {Doppmann}, {Covey}, \&
  {Hillenbrand}}]{White2007}
{White}, R.~J., {Greene}, T.~P., {Doppmann}, G.~W., {Covey}, K.~R., \&
  {Hillenbrand}, L.~A. 2007, in Protostars and Planets V, ed. B.~{Reipurth},
  D.~{Jewitt}, \& K.~{Keil}, 117, \dodoi{10.48550/arXiv.astro-ph/0604081}

\bibitem[{{Wilking} {et~al.}(2008){Wilking}, {Gagn{\'e}}, \&
  {Allen}}]{Wilking2008}
{Wilking}, B.~A., {Gagn{\'e}}, M., \& {Allen}, L.~E. 2008, in Handbook of Star
  Forming Regions, ed. B.~{Reipurth}, Vol.~5 (Monograph Publications), 351,
  \dodoi{10.48550/arXiv.0811.0005}

\bibitem[{{Winn} {et~al.}(2010){Winn}, {Fabrycky}, {Albrecht}, \&
  {Johnson}}]{Winn2010}
{Winn}, J.~N., {Fabrycky}, D., {Albrecht}, S., \& {Johnson}, J.~A. 2010, \apjl,
  718, L145, \dodoi{10.1088/2041-8205/718/2/L145}

\bibitem[{{Winn} \& {Fabrycky}(2015)}]{Winn2015}
{Winn}, J.~N., \& {Fabrycky}, D.~C. 2015, \araa, 53, 409,
  \dodoi{10.1146/annurev-astro-082214-122246}

\bibitem[{{Winter} {et~al.}(2024){Winter}, {Benisty}, {Shuai}, {D{\^u}chene},
  {Cuello}, {Anania}, {Cadiou}, \& {Joncour}}]{Winter2024}
{Winter}, A.~J., {Benisty}, M., {Shuai}, L., {et~al.} 2024, Astronomy and
  Astrophysics, 691, A43, \dodoi{10.1051/0004-6361/202450842}

\bibitem[{{Wright} {et~al.}(2004){Wright}, {Marcy}, {Butler}, \&
  {Vogt}}]{Wright2004}
{Wright}, J.~T., {Marcy}, G.~W., {Butler}, R.~P., \& {Vogt}, S.~S. 2004, The
  Astrophysical Journal Supplement Series, 152, 261, \dodoi{10.1086/386283}

\bibitem[{{Wu} \& {Lithwick}(2011)}]{Wu2011}
{Wu}, Y., \& {Lithwick}, Y. 2011, The Astrophysical Journal, 735, 109,
  \dodoi{10.1088/0004-637X/735/2/109}

\bibitem[{{Wyatt}(2008)}]{Wyatt2008}
{Wyatt}, M.~C. 2008, Annual Review of Astronomy and Astrophysics, 46, 339,
  \dodoi{10.1146/annurev.astro.45.051806.110525}

\bibitem[{{Wyatt} {et~al.}(2012){Wyatt}, {Kennedy}, {Sibthorpe},
  {Moro-Mart{\'\i}n}, {Lestrade}, {Ivison}, {Matthews}, {Udry}, {Greaves},
  {Kalas}, {Lawler}, {Su}, {Rieke}, {Booth}, {Bryden}, {Horner}, {Kavelaars},
  \& {Wilner}}]{Wyatt2012}
{Wyatt}, M.~C., {Kennedy}, G., {Sibthorpe}, B., {et~al.} 2012, Monthly Notices
  of the Royal Astronomical Society, 424, 1206,
  \dodoi{10.1111/j.1365-2966.2012.21298.x}

\bibitem[{{Yang} {et~al.}(2023){Yang}, {Zhang}, {Meng}, {Han}, {Misra}, {Yang},
  \& {Pi}}]{Yang2023}
{Yang}, Z., {Zhang}, L., {Meng}, G., {et~al.} 2023, Astronomy and Astrophysics,
  669, A15, \dodoi{10.1051/0004-6361/202142710}

\bibitem[{{Yu} {et~al.}(2024){Yu}, {Aigrain}, {Klein}, {Barrag{\'a}n},
  {Mortier}, {O'Sullivan}, \& {Cretignier}}]{Yu2024}
{Yu}, H., {Aigrain}, S., {Klein}, B., {et~al.} 2024, Monthly Notices of the
  Royal Astronomical Society, 528, 5511, \dodoi{10.1093/mnras/stae137}

\bibitem[{{Zakhozhay} {et~al.}(2022){Zakhozhay}, {Launhardt}, {M{\"u}ller},
  {Brems}, {Eigenthaler}, {Gennaro}, {Hempel}, {Hempel}, {Henning}, {Kennedy},
  {Kim}, {K{\"u}rster}, {Lachaume}, {Manerikar}, {Patel}, {Pavlov}, {Reffert},
  \& {Trifonov}}]{Zakhozhay2022}
{Zakhozhay}, O.~V., {Launhardt}, R., {M{\"u}ller}, A., {et~al.} 2022, Astronomy
  and Astrophysics, 667, A63, \dodoi{10.1051/0004-6361/202244213}

\bibitem[{{Z{\'u}{\~n}iga-Fern{\'a}ndez}
  {et~al.}(2021){Z{\'u}{\~n}iga-Fern{\'a}ndez}, {Bayo}, {Elliott}, {Zamora},
  {Corval{\'a}n}, {Haubois}, {Corral-Santana}, {Olofsson}, {Hu{\'e}lamo},
  {Sterzik}, {Torres}, {Quast}, \& {Melo}}]{Zuniga-Fernandez2021}
{Z{\'u}{\~n}iga-Fern{\'a}ndez}, S., {Bayo}, A., {Elliott}, P., {et~al.} 2021,
  \aap, 645, A30, \dodoi{10.1051/0004-6361/202037830}

\end{thebibliography}
\bibliographystyle{aasjournal}



\appendix \phantomsection \label{sec:Appendix}

\FloatBarrier
\section{Updated Inclination Measurements of Debris Disk Systems}\label{appendix:Distributions}

Here, we provide \idisk, \istar, and \Deltai distributions for debris disk systems (Figure~\ref{fig:sample_posteriors_1}). The top panel in the column shows the inferred probability distribution of the measured disk inclination \idisk (see Section \ref{subsec:PPD_Sample}). The middle panel displays the posterior distribution of the inclination of the star's rotation axis \istar (see Section \ref{subsec:Stellar_Inclinations}). The lower panel gives the probability distribution of the minimum star-disk obliquity \Deltai, computed by sampling the distributions of \idisk and \istar (see Section \ref{subsec:Star-Disk_Obliquity}).

% \clearpage

\begin{figure}[h!]
    \centering
    \begin{minipage}[h!]{\textwidth}
        \centering
        \includegraphics[width=0.94\textwidth]{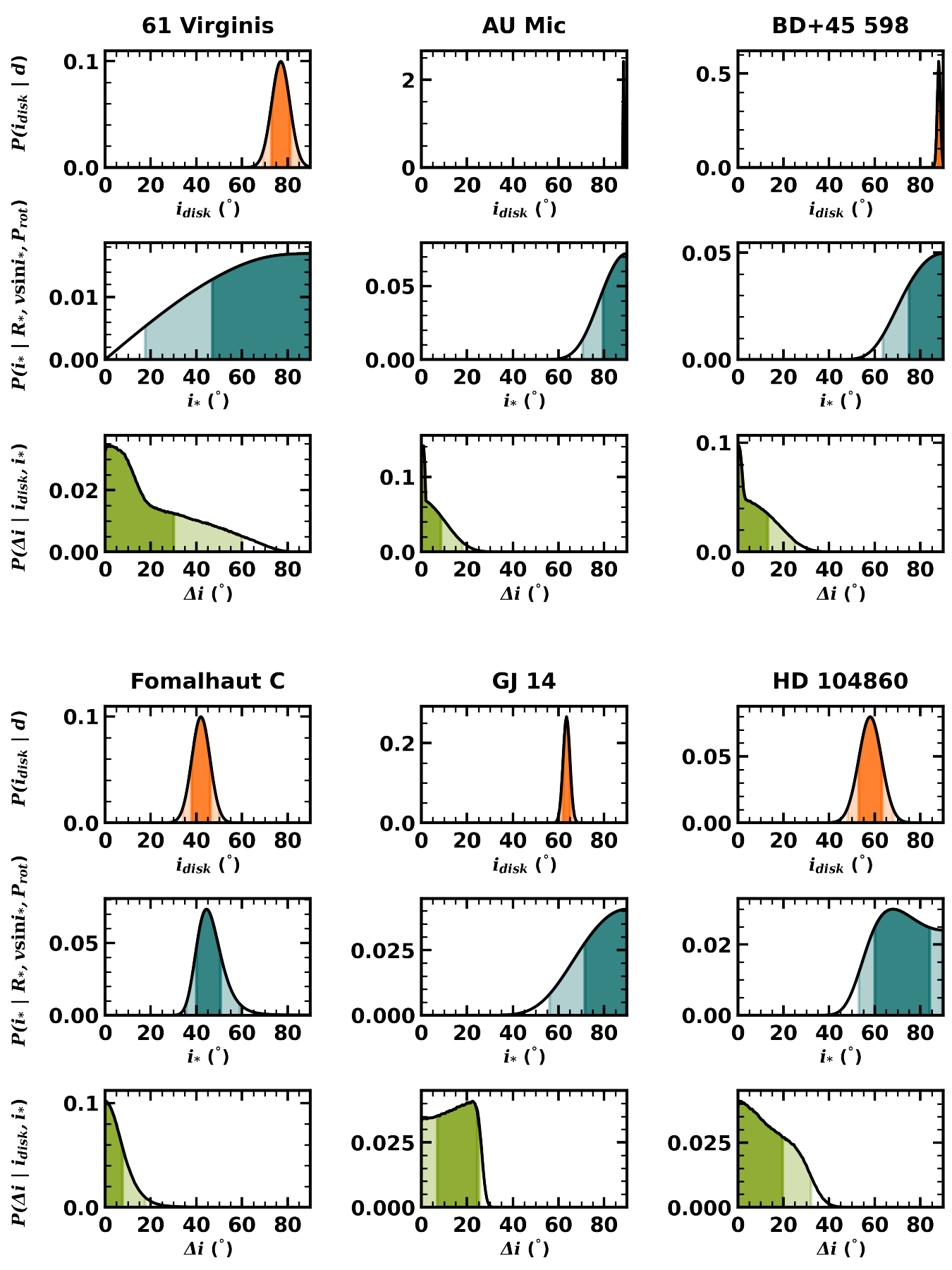}
        \caption{The probability distribution functions of $i_{\textrm{disk}}$, $i_{*}$, and $\Delta i$ for debris disk systems. Dark and light shaded regions indicate the 1$\sigma$ and 2$\sigma$ highest density intervals, respectively.}
    	\label{fig:sample_posteriors_1}
    \end{minipage}
\end{figure}
\addtocounter{figure}{-1}

\begin{figure}[h!]
    \centering
    \begin{minipage}[h!]{\textwidth}
        \centering
        \includegraphics[width=0.94\textwidth]{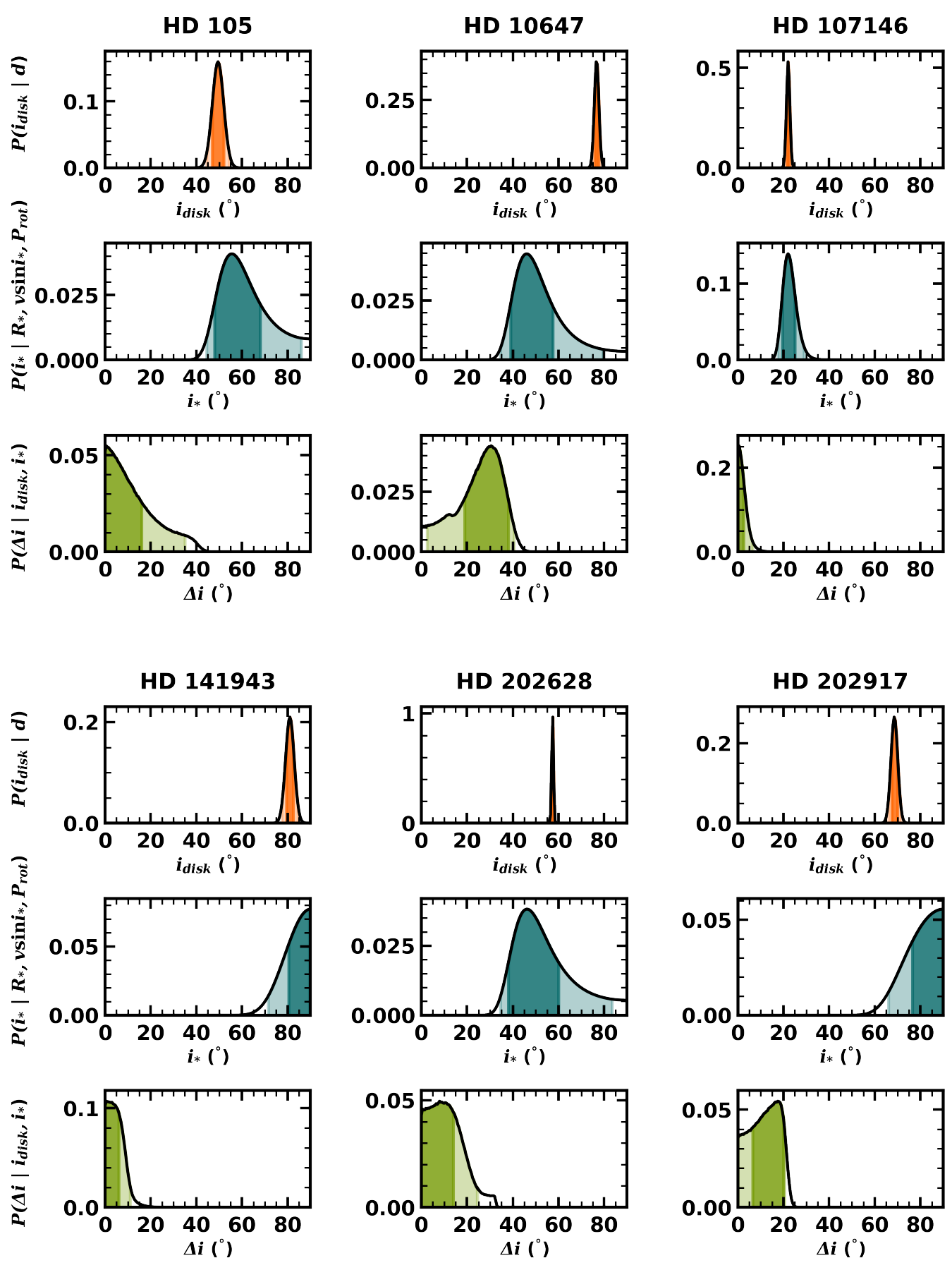}
        \caption{\textit{Continued.}}
    	\label{fig:sample_posteriors_2}
    \end{minipage}
\end{figure}
\addtocounter{figure}{-1}

\begin{figure}[h!]
    \centering
    \begin{minipage}[h!]{\textwidth}
        \centering
        \includegraphics[width=0.94\textwidth]{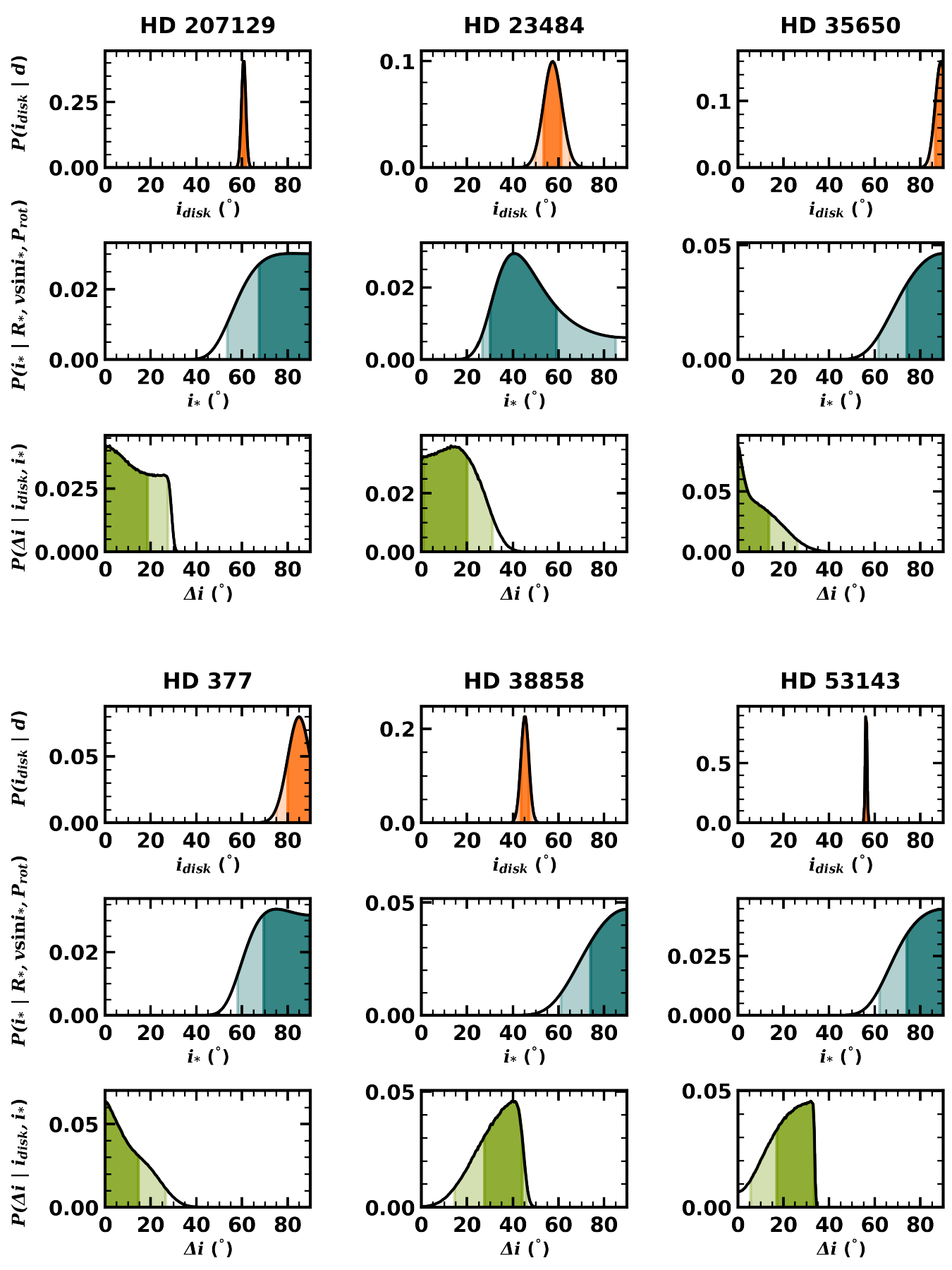}
        \caption{\textit{Continued.}}
    	\label{fig:sample_posteriors_3}
    \end{minipage}
\end{figure}
\addtocounter{figure}{-1}

\begin{figure}[h!]
    \centering
    \begin{minipage}[h!]{\textwidth}
        \centering
        \includegraphics[width=0.94\textwidth]{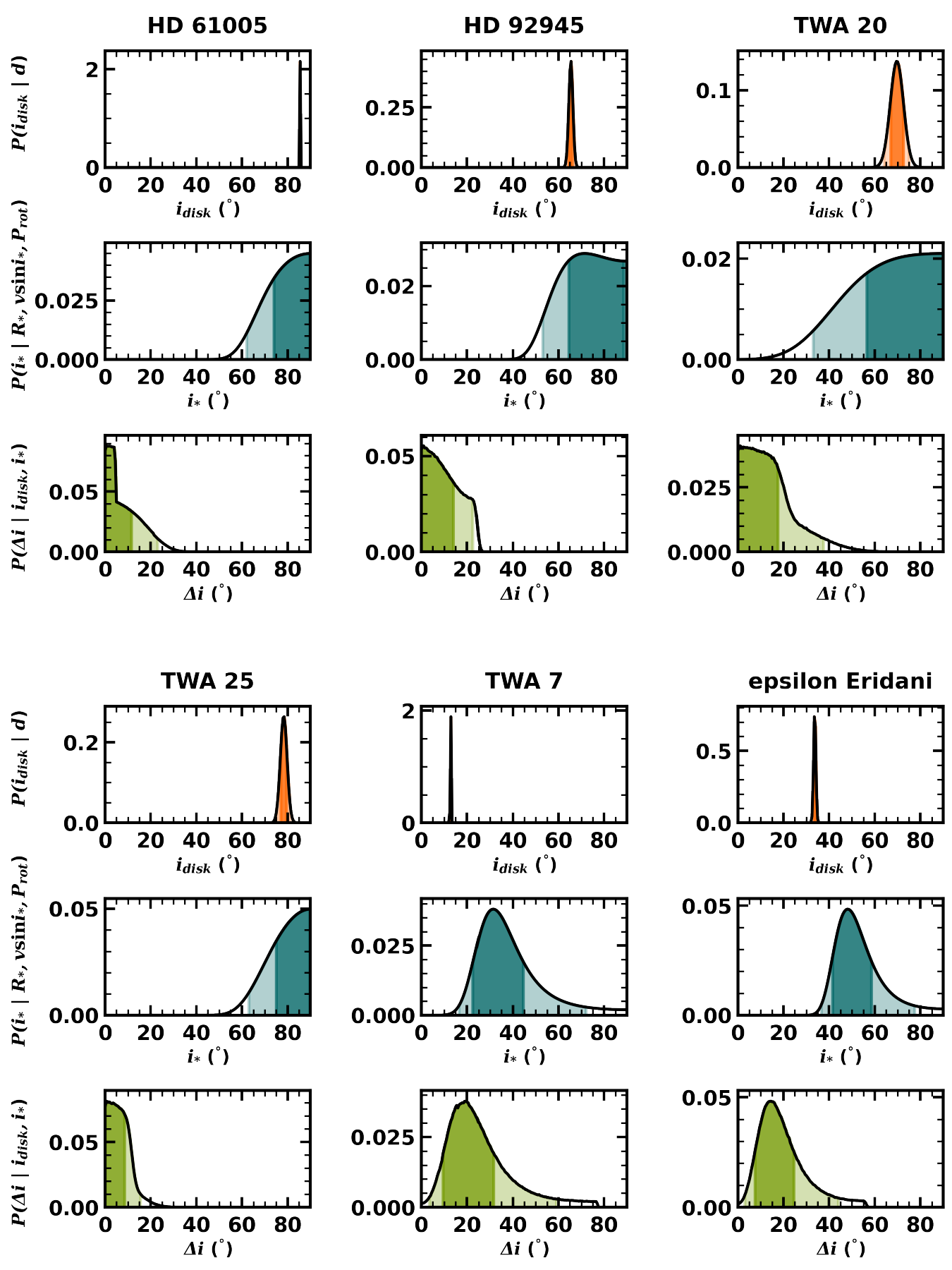}
        \caption{\textit{Continued.}}
    	\label{fig:sample_posteriors_4}
    \end{minipage}
\end{figure}

\clearpage

\section{Compiled and Adopted Sample Properties}\label{appendix:Tables}
In this section we tabulate the adopted values and results for each object Table \ref{tab:Sample_Properties}. We also provide a compilation of stellar and disk properties from the literature that we use in this analysis (Section \ref{sec:Sample_Descriptions}). The adopted values are highlighted in bold text in the last row associated with the object. Stellar spectral type and \Teff is provided in Table~\ref{tab:DD_Sample_List}. Published values of stellar \vsini, \Prot, and \Rstar used to compute the weighted mean are provided in Table~\ref{tab:DD_Rstar_Prot_vsini}. Star and debris disk inclinations are displayed in Table~\ref{tab:DD_Disk_Inclinations}.

\setlength{\LTcapwidth}{0.95\textwidth}

\begin{center}
\clearpage
\begin{longtable}{llllllllll}
\caption{Summary of relevant system properties of all samples. Protoplanetary disk properties are from \citet{Biddle2025}. The process of obtaining debris disks properties is described in Section \ref{subsec:DD_Sample} in this work, with all measurements in the literature that were used to compute the weighted mean of each measurement tabulated in Tables \ref{tab:DD_Sample_List}, \ref{tab:DD_Rstar_Prot_vsini}, and \ref{tab:DD_Disk_Inclinations}. Warm and hot Jupiter properties listed below are adopted from \citet{Morgan2024}. For transiting planets, the inclination of the orbital plane, \iorb, is assumed to be 90$^{\circ}$.}
\label{tab:Sample_Properties}
\\
\hline
Identifier & Age   & sep  & \Teff &  \Rstar  &  \vsini	      & \Prot &  \istar       & \iorb        & \Deltai      \\
           & (Myr) & (au) & (K)   &  (\Rsun) &  (km~s$^{-1}$) & (d)   &  ($^{\circ}$) & ($^{\circ}$) & ($^{\circ}$) \\
\hline
\endfirsthead
\hline
Identifier & Age   & sep  & \Teff &  \Rstar  &  \vsini	      & \Prot &  \istar       & \iorb        & \Deltai      \\
           & (Myr) & (au) & (K)   &  (\Rsun) &  (km~s$^{-1}$) & (d)   &  ($^{\circ}$) & ($^{\circ}$) & ($^{\circ}$) \\
\hline
\endhead
\multicolumn{10}{c}{Protoplanetary Disks} \\
\hline
2M J0412 & 4.3$\pm$1.9 & 126.0$^{+0.9}_{-0.6}$ & 3337$\pm$3 & 1.0$\pm$0.2 & 6.4$\pm$0.7 & 5.7$\pm$0.2 & 48.2$^{+21.4}_{-10.8}$ & 15.8$\pm$0.8 & 34.2$^{+19.5}_{-12.7}$ \\
2M J0420 & 6.6$\pm$0.1 & 21.1$\pm$0.0 & 3083$\pm$32 & 0.9$\pm$0.4 & 14.2$\pm$0.8 & 2.2$\pm$0.1 & 47.3$^{+29.8}_{-11.2}$ & 38.2$\pm$0.2 & 0.2$^{+30.5}_{-0.2}$ \\
2M J0433 & 3.0$\pm$2.8 & 46.2$^{+0.1}_{-0.1}$ & 3142$\pm$4 & 1.5$\pm$0.4 & 16.0$\pm$0.8 & 5.9$\pm$0.2 & 90.0$^{+0.0}_{-21.8}$ & 57.6$\pm$0.1 & 32.1$^{+0.2}_{-21.1}$ \\
2M J0434 & 7.2$\pm$3.0 & 81.2$\pm$0.8 & 3293$\pm$3 & 0.9$\pm$0.2 & 8.7$\pm$0.6 & 6.0$\pm$0.2 & 90.0$^{+0.0}_{-18.9}$ & 68.5$\pm$0.2 & 0.2$^{+13.8}_{-0.2}$ \\
2M J0436 & 2.5$\pm$2.5 & 13.2$^{+0.4}_{-0.3}$ & 3592$\pm$85 & 1.0$\pm$0.1 & 16.4$\pm$0.8 & 2.7$\pm$0.1 & 63.6$^{+15.1}_{-8.4}$ & 53.4$\pm$1.4 & 7.1$^{+14.0}_{-7.1}$ \\
2M J1100 & 3.5$\pm$0.7 & 99.1$^{+15.2}_{-13.3}$ & 3340$\pm$64 & 0.9$\pm$0.2 & 5.4$\pm$0.8 & 2.8$\pm$0.1 & 19.6$^{+6.5}_{-4.5}$ & 4.4$\pm$3.1 & 15.0$^{+7.3}_{-5.2}$ \\
2M J1608 & 2.5$\pm$0.3 & 100.8$^{+4.7}_{-3.1}$ & 4856$\pm$53 & 1.7$\pm$0.3 & 8.0$\pm$0.0 & 6.2$\pm$0.2 & 28.0$^{+11.2}_{-14.5}$ & 72.5$\pm$0.6 & 45.5$^{+13.2}_{-12.4}$ \\
\hline
\end{longtable}
\end{center}

\clearpage

\begin{table}
\centering
\caption{List of debris disk systems, coordinates, and compiled values of \Teff, with the adopted weighted mean shown in bold. }
\label{tab:DD_Sample_List}
\begin{tabular}{lllll}
\\
\toprule
Identifier  &	RA (J2000)    &	Dec (J2000)  &  \Teff  &  Reference \\
            &	(h m s)       &	($^{\circ}$ ' ")  &  (K)     &	          \\
\midrule
61 Vir  &  13 18 24.3140  &  -18 18 40.2977  &  5562$\pm$109  &  \citet{Stassun2019}  \\
$\cdots$  &  $\cdots$  &  $\cdots$  &  5755$\pm$58  &  \citet{Borisov2023}  \\
$\cdots$  &  $\cdots$  &  $\cdots$  &  5572$\pm$50  &  \citet{Perdelwitz2024}  \\
$\cdots$  &  $\cdots$  &  $\cdots$  &  5578  &  \citet{Luck2017}  \\
$\cdots$  &  $\cdots$  &  $\cdots$  &  5533$\pm$19  &  \citet{Gonzalez2010}  \\
\end{tabular}
\end{table}
\addtocounter{table}{-1}

\begin{table}
\centering
\caption{\textit{Continued.}}
\begin{tabular}{lllll}
\\
\toprule
Identifier  &	RA (J2000)    &	Dec (J2000)  &  \Teff  &  Reference \\
            &	(h m s)       &	($^{\circ}$ ' ")  &  (K)     &	          \\
\midrule
HD 104860  &  12 04 33.7309  &  +66 20 11.7128  &  6000$\pm$146  &  \citet{Stassun2019}  \\
$\cdots$  &  $\cdots$  &  $\cdots$  &  5873$\pm$364  &  \citet{Sarmento2020}  \\
$\cdots$  &  $\cdots$  &  $\cdots$  &  5935$\pm$85  &  \citet{Abdurro'uf2022}  \\
$\cdots$  &  $\cdots$  &  $\cdots$  &  \textbf{5948$\pm$72}  &  \textbf{Adopted}  \\
\midrule									
HD 105  &  00 05 52.5448  &  -41 45 11.0450  &  6033$\pm$144  &  \citet{Stassun2019}  \\
\end{tabular}
\end{table}
\addtocounter{table}{-1}

\begin{table}
\centering
\caption{\textit{Continued.}}
\begin{tabular}{lllll}
\\
\toprule
Identifier  &	RA (J2000)    &	Dec (J2000)  &  \Teff  &  Reference \\
            &	(h m s)       &	($^{\circ}$ ' ")  &  (K)     &	          \\
\midrule                              
HD 202628  &  21 18 27.2696  &  -43 20 04.7430  &  5831$\pm$133  &  \citet{Stassun2019}  \\
$\cdots$  &  $\cdots$  &  $\cdots$  &  5816  &  \citet{Soto2018}  \\
$\cdots$  &  $\cdots$  &  $\cdots$  &  5833$\pm$11  &  \citet{dosSantos2016}  \\
$\cdots$  &  $\cdots$  &  $\cdots$  &  5831$\pm$13  &  \citet{Hojjatpanah2020}  \\
$\cdots$  &  $\cdots$  &  $\cdots$  &  \textbf{5832$\pm$8}  &  \textbf{Adopted}  \\
\end{tabular}
\end{table}
\addtocounter{table}{-1}

\begin{table}
\centering
\caption{\textit{Continued.}}
\begin{tabular}{lllll}
\\
\toprule
Identifier  &	RA (J2000)    &	Dec (J2000)  &  \Teff  &  Reference \\
            &	(h m s)       &	($^{\circ}$ ' ")  &  (K)     &	          \\
\midrule 								
TWA 7  &  10 42 30.1019  &  -33 40 16.2298  &  3509$\pm$120  &  \citet{Stassun2019}  \\
$\cdots$  &  $\cdots$  &  $\cdots$  &  3541$\pm$10  &  \citet{Loyd2014}  \\
$\cdots$  &  $\cdots$  &  $\cdots$  &  \textbf{3541$\pm$10}  &  \textbf{Adopted}  \\
\midrule			
TWA 20 &	12 31 38.0692   &	-45 58 59.4579  &  3504$\pm$120  &  \citet{Stassun2019}  \\
$\cdots$  &  $\cdots$  &  $\cdots$  &  \textbf{3504$\pm$120}  &  \textbf{Adopted}  \\
\midrule	
TWA 25  &  12 15 30.7203  &  +39 48 42.6067  &  3899$\pm$120  &  \citet{Stassun2019}  \\
$\cdots$  &  $\cdots$  &  $\cdots$  &  3742  &  \citet{daSilva2009}  \\
$\cdots$  &  $\cdots$  &  $\cdots$  &  \textbf{3899$\pm$120}  &  \textbf{Adopted}  \\
\botrule
\end{tabular}
\end{table}

\begin{table}
\centering
\caption{Compilation of literature and adopted values of \Rstar, \Prot, and \vsini for debris disk systems.}
\label{tab:DD_Rstar_Prot_vsini}
\begin{tabular}{p{1.8cm}p{1.6cm}p{2cm}p{1.3cm}p{3.0cm}p{1.6cm}p{4.4cm}} \\
\toprule
Identifier & \Rstar & Reference & \Prot & Reference & \vsini & Reference \\
 & (\Rsun) &  & (d) &  & (km s$^{-1}$) &  \\
\midrule
61 Vir  &  0.984$\pm$0.045  &  \citet{Stassun2019}  &  29.0$\pm$2.9  &  \citet{Baliunas1996}  &  $<$2.0  &  \citet{Nordstrom2004}  \\
$\cdots$  &  $\cdots$  &  $\cdots$  &  33  &  \citet{Wright2004}  &  $\cdots$  &  $\cdots$  \\
$\cdots$  &  $\cdots$  &  $\cdots$  &  32.1$\pm$0.2  &  \citet{Yu2024}  &  $\cdots$  &  $\cdots$  \\
$\cdots$  &  \textbf{0.984$\pm$0.082}  &  \textbf{\,\,Adopted}  &  \textbf{32.1$\pm$4.2}  &  \textbf{Adopted}  &  \textbf{$<$2.0}  &  \textbf{Adopted}  \\
\midrule													
AU Mic  &  0.698$\pm$0.021  &  \citet{Stassun2019}  &  4.87$\pm$0.31  &  This Work  &  9.5$\pm$0.2  &  \citet{Hojjatpanah2019}  \\
\end{tabular}
\end{table}
\addtocounter{table}{-1}

\begin{table}
\centering
\caption{\textit{Continued.}}
\begin{tabular}{p{1.8cm}p{1.6cm}p{2cm}p{1.3cm}p{3.0cm}p{1.6cm}p{4.4cm}}\\
\toprule
Identifier  &  \Rstar  &  Reference  &  \Prot  &  Reference  &  \vsini	      &	Reference  \\
	          & (\Rsun)	  &		          &     (d)	  &		          &	(km s$^{-1}$)	&             \\
\midrule	
$\cdots$  &  $\cdots$  &  $\cdots$  &  $\cdots$  &  $\cdots$  &  3.2$\pm$0.4  &  \citet{Gonzalez2010}  \\
$\cdots$  &  $\cdots$  &  $\cdots$  &  $\cdots$  &  $\cdots$  &  2.5$\pm$0.5  &  \citet{Gonzalez2010}  \\
$\cdots$  &  $\cdots$  &  $\cdots$  &  $\cdots$  &  $\cdots$  &  8  &  \citet{Takeda2005}  \\
$\cdots$  &  $\cdots$  &  $\cdots$  &  $\cdots$  &  $\cdots$  &  4.73  &  \citet{Abdurro'uf2022}  \\
$\cdots$  &  $\cdots$  &  $\cdots$  &  $\cdots$  &  $\cdots$  &  2.9$\pm$0.5  &  \citet{Brewer2015}  \\
\end{tabular}
\end{table}
\addtocounter{table}{-1}

\begin{table}
\centering
\caption{\textit{Continued.}}
\begin{tabular}{p{1.8cm}p{1.6cm}p{2cm}p{1.3cm}p{3.0cm}p{1.6cm}p{4.4cm}}\\
\toprule
Identifier  &  \Rstar  &  Reference  &  \Prot  &  Reference  &  \vsini	      &	Reference  \\
	          & (\Rsun)	  &		          &     (d)	  &		          &	(km s$^{-1}$)	&             \\
\midrule	
HD 10647  &  1.096$\pm$0.057  &  \citet{Stassun2019}  &  8.90$\pm$0.66  &  This Work  &  3.8$\pm$0.4  &  \citet{Soto2018}  \\
$\cdots$  &  $\cdots$  &  $\cdots$  &  $\cdots$  &  $\cdots$  &  5.61$\pm$0.5  &  \citet{Valenti2005}  \\
$\cdots$  &  $\cdots$  &  $\cdots$  &  $\cdots$  &  $\cdots$  &  5.82$\pm$1.95  &  \citet{Borisov2023}  \\
$\cdots$  &  $\cdots$  &  $\cdots$  &  $\cdots$  &  $\cdots$  &  5.2  &  \citet{Schroder2009}  \\
$\cdots$  &  $\cdots$  &  $\cdots$  &  $\cdots$  &  $\cdots$  &  3.81$\pm$0.40  &  \citet{Soto2018}  \\
\end{tabular}
\end{table}
\addtocounter{table}{-1}

\begin{table}
\centering
\caption{\textit{Continued.}}
\begin{tabular}{p{1.8cm}p{1.6cm}p{2cm}p{1.3cm}p{3.0cm}p{1.6cm}p{4.4cm}}\\
\toprule
Identifier  &  \Rstar  &  Reference  &  \Prot  &  Reference  &  \vsini	      &	Reference  \\
	          & (\Rsun)	  &		          &     (d)	  &		          &	(km s$^{-1}$)	&             \\
\midrule												
HD 23484  &  0.804$\pm$0.054  &  \citet{Stassun2019}  &  9.75$\pm$1.29  &  This Work  &  3.1$\pm$0.7  &  \citet{Fuhrmann2017}  \\
$\cdots$  &  $\cdots$  &  $\cdots$  &  $\cdots$  &  $\cdots$  &  2.4$\pm$0.5  &  \citet{Valenti2005}  \\
$\cdots$  &  $\cdots$  &  $\cdots$  &  $\cdots$  &  $\cdots$  &  2.9$\pm$1.2  &  \citet{Torres2006}  \\
$\cdots$  &  \textbf{0.804$\pm$0.078}  &  \textbf{\,\,Adopted}  &  \textbf{9.8$\pm$1.4}  &  \textbf{Adopted}  &  \textbf{2.7$\pm$0.4}  &  \textbf{Adopted}  \\
\midrule													
HD 35650  &  0.672$\pm$0.063  &  \citet{Stassun2019}  &  9.65$\pm$0.54  &  This Work  &  4.0$\pm$0.2  &  \citet{Torres2006}  \\
\end{tabular}
\end{table}
\addtocounter{table}{-1}

\begin{table}
\centering
\caption{\textit{Continued.}}
\begin{tabular}{p{1.8cm}p{1.6cm}p{2cm}p{1.3cm}p{3.0cm}p{1.6cm}p{4.4cm}}\\
\toprule
Identifier  &  \Rstar  &  Reference  &  \Prot  &  Reference  &  \vsini	      &	Reference  \\
	          & (\Rsun)	  &		          &     (d)	  &		          &	(km s$^{-1}$)	&             \\
\midrule
HD 92945  &  0.791$\pm$0.056  &  \citet{Stassun2019}  &  7.47$\pm$0.53  &  This Work  &  5.41$\pm$0.21  &  \citet{Perdelwitz2024}  \\
$\cdots$  &  $\cdots$  &  $\cdots$  &  $\cdots$  &  $\cdots$  &  5.1$\pm$0.5  &  \citet{Valenti2005}  \\
$\cdots$  &  $\cdots$  &  $\cdots$  &  $\cdots$  &  $\cdots$  &  4.0$\pm$0.5  &  \citet{Nordstrom2004}  \\
$\cdots$  &  $\cdots$  &  $\cdots$  &  $\cdots$  &  $\cdots$  &  6.1$\pm$3.0  &  \citet{Herrero2012}  \\
$\cdots$  &  $\cdots$  &  $\cdots$  &  $\cdots$  &  $\cdots$  &  6.1$\pm$3.0  &  \citet{Herrero2012}  \\
\botrule
\end{tabular}
\end{table}

\begin{table}
\centering
\caption{Literature and adopted values of debris disk inclinations. Asymmetric \idisk measurements were converted to symmetric measurements by adopting the average of the upper and lower bounds on the original uncertainties.}
\label{tab:DD_Disk_Inclinations}
\begin{tabular}{lll}
\\
\toprule
Identifier  &  \idisk	  &  Reference  \\
	        &	($^{\circ}$)     &	           \\
\midrule
61 Vir$^{a}$  &  77.0$\pm$4.0  &  \citet{Wyatt2012}  \\
$\cdots$  &  \textbf{77.0$\pm$4.0}  &  \textbf{Adopted}  \\
\midrule					
AU Mic$^{a}$  &  88.4$\pm$0.1  &  \citet{Matra2025}  \\
$\cdots$  &  \textbf{88.4$\pm$0.1}  &  \textbf{Adopted}  \\
\midrule					
BD+45 598  &  88.1$\pm$0.7  &  \citet{Vincent2024}  \\
\end{tabular}
\end{table}
\addtocounter{table}{-1}

\begin{table}
\centering
\caption{Table Caption. }
\begin{tabular}{lll}
\\
\toprule
Identifier  &  \idisk  &  Reference  \\
	        &	($^{\circ}$)     &	           \\
\midrule
HD 377  &  85$\pm$5  &  \citet{Choquet2016}  \\
$\cdots$  &  \textbf{85.0$\pm$5.0}  &  \textbf{Adopted}  \\
\midrule	
HD 38858  &  45.9$^{+1.1}_{-2.4}$  &  \citet{Marshall2021}  \\
$\cdots$  &  \textbf{45.3$\pm$1.8}  &  \textbf{Adopted}  \\
\midrule					
HD 53143  &  56.23$\pm$0.37  &  \citet{MacGregor2022}  \\
$\cdots$  &  \textbf{56.23$\pm$0.37}  &  \textbf{Adopted}  \\
\midrule					
HD 61005  &  85.6$\pm$0.1  &  \citet{Crotts2024}  \\
$\cdots$  &  \textbf{85.6$\pm$0.1}  &  \textbf{Adopted}  \\
\midrule					
HD 92945  &  65.4$\pm$0.9  &  \citet{Marino2019}  \\
$\cdots$  &  \textbf{65.4$\pm$0.9}  &  \textbf{Adopted}  \\
\midrule				
TWA 7$^{a}$  &  12.91$^{+0.18}_{-0.19}$  &  \citet{Ren2021}  \\
$\cdots$  &  \textbf{12.9$\pm$0.2}  &  \textbf{Adopted}  \\
\midrule			
TWA 20  &  70.1$^{+2.5}_{-3.3}$  &  \citet{Palatnick2025}  \\
$\cdots$  &  \textbf{69.7$\pm$2.9}  &  \textbf{Adopted}  \\
\midrule	
TWA 25  &  75.0$\pm$6.0  &  \citet{Choquet2016}  \\
$\cdots$  &  \textbf{75.0$\pm$6.0}  &  \textbf{Adopted}  \\
\botrule
\multicolumn{3}{c}{$^{a}$Updated from the original \citet{Hurt2023} catalog.}
\end{tabular}
\end{table}

\end{document}